\documentclass{aa}
\usepackage{graphicx}
\usepackage{xcolor}
\usepackage{txfonts}
\usepackage{booktabs}
\usepackage{multirow}
\usepackage{makecell}
\usepackage{nicefrac}
\usepackage{siunitx}
\usepackage{longtable}
\usepackage{subfig}
\usepackage{siunitx}
\usepackage[colorlinks=true,linkcolor=blue,allcolors=blue]{hyperref}

\newcommand{\Halpha}{{\text{H}\ensuremath{\alpha}}}
\newcommand{\Hbeta}{{\text{H}\ensuremath{\beta}}}

\let\unit\si

\begin{document}

   \title{A new sample of Little Red Dots at $z<0.45$ in DESI DR1}
   \subtitle{Broad Balmer lines, low ionization spectrum and no variability}
   \titlerunning{DESI LRDs}

    \author{
    Kevin~Park\inst{1}\thanks{E-mail: Kevin.Park@ista.ac.at}
    \and Alberto~Torralba\inst{1}
    \and Jorryt~Matthee\inst{1}
    \and Sara~Mascia\inst{1}
    \and Zolt\'an~Haiman\inst{1,2,3}
    \and Rohan~P.~Naidu\inst{4}\thanks{NASA Hubble Fellow, Pappalardo Fellow}
    \and Anna~de~Graaff\inst{5, 6}\thanks{Clay Fellow}
    }
    
    \institute{
    Institute of Science and Technology Austria (ISTA), Am Campus 1, 3400 Klosterneuburg, Austria
    \and
    Department of Astronomy, Columbia University, 550 W 120th St, New York, NY 10027, USA
    \and
    Department of Physics, Columbia University, 550 W 120th St, New York, NY 10027, USA
    \and
    MIT Kavli Institute for Astrophysics and Space Research, 70 Vassar Street, Cambridge, MA 02139, USA
    \and
    Center for Astrophysics, Harvard \& Smithsonian, 60 Garden St, Cambridge, MA 02138, USA
    \and 
    Max-Planck-Institut f\"ur Astronomie, K\"onigstuhl 17, D-69117 Heidelberg, Germany
    }

  \date{Accepted XXX. Received YYY; in original form ZZZ}

  \abstract{
JWST has unveiled an abundant population of compact broad-line emitters largely at $z\gtrsim4$, the Little Red Dots (LRDs), which might represent a previously unprobed supermassive black hole evolution channel predominant at high redshift. However, the LRDs have remained mostly elusive at lower redshift ($z\lesssim2$) where detailed studies are possible from ground-based observatories. We searched for low-redshift LRDs in the Dark Energy Spectroscopic Instrument (DESI) survey. Our search is primarily based on emission line properties, as opposed to earlier approaches that searched for compact sources with specific photometric spectral energy distributions. We report the discovery of eight LRDs at $z=0.2-0.45$, which show spectral features akin to the high-redshift LRDs in the rest-frame optical. The sources are characterized by broad Balmer lines, steep Balmer decrements, compact morphologies, Balmer absorption features and/or strong He\,\textsc{i} emission, but weak or absent He\,\textsc{ii}, [Ne\,\textsc{v}] or other high excitation lines typical of Type I AGN. For 7 out of 8 sources, we retrieve dense-cadence light curves from time-domain surveys and for most sources we find weak to no intrinsic variability ($0.0-0.1$ mag) over 4--17 years in the rest-frame. We also highlight the identification of a quasar with  similar Balmer line profiles as LRDs, but that shows differences in Balmer decrement, significant variability, and high-ionisation lines. Given the effective volume $4.9 \; {\rm Gpc^3}$ covered by DESI DR1 at $z<0.45$, our sample corresponds to a number density of $1.6\times10^{-9}$Mpc$^{-3}$, indicating a number density $\sim$10,000 times lower than in the first billion years of cosmic time. We find a dearth of luminous and red LRDs at $z<1$ compared to higher-redshift, which could suggest lower gas feeding rates of LRD activity due to higher metallicities at later cosmic epochs.
  }


   \maketitle

\section{Introduction}\label{sec:introduction}

One of the most remarkable discoveries of the James Webb Space Telescope (JWST) has been the identification of an abundant population of faint broad Balmer line emitters with compact morphologies $\lesssim 100$ pc at high-redshift $z\sim3$--$9$, nicknamed the "Little Red Dots" \citep[LRDs;][]{matthee2024a}. This population is characterized by red UV to optical colors in the spectral energy distribution (SED) and, in most cases, a blue UV slope, usually described as a ``V-shape'' \citep[e.g.,][]{kocevski2023, kokorev2024a, matthee2024a, labbe2025} with an inflection point near the Balmer limit $\sim 3645\AA$ \citep[e.g.,][]{setton2024, Hviding2025}.

Due to their compactness and their broad Balmer lines, LRDs are mainly interpreted as being powered by accreting supermassive black holes (SMBH). However, the typical models of active galactic nuclei (AGN) involving a hot accretion disk, broad line region (BLR), and dusty torus struggle to reproduce the peculiarities of LRDs.
Unlike most AGNs, LRDs are typically found to be faint in X-rays \citep[e.g.,][]{yue2024, ananna2024, Kokubo25, maiolino2025}, they lack hot and warm mid-IR dust emission \citep[e.g.,][]{leung2024, Setton2025, Xiao25, Casey25, delvecchio_active_2025}, as well as radio emission \citep[][]{latif2025, perger2025, mazzolari2024}.
Beyond the V-shape, most LRDs show a combination of spectral features that are unusual in typical AGN. In the first place, they show very high Balmer decrements \citep[up to $\Halpha/\Hbeta\sim10$; e.g.,][]{Barro2025_cliff_virgil, degraaff25BH*, Matthee2026,PerezGonzalez26, Sun26} that cannot be simply explained by dust attenuation \citep[e.g.,][]{nikopoulos2025, Torralba2026_gn9771}, and extremely high \Halpha{} equivalent widths \citep[up to $\gtrsim 1000$~\AA{}]{degraaff25BH*, Matthee2026}. Moreover the \Halpha{} broad wings are remarkably symmetric, and well described by exponential profiles suggesting a broadening by electron scattering and other radiative-transfer effects \citep[e.g.,][although cf. \citealt{brazzini2025, Scholtz2026}]{rusakov2025, Chang25, naidu2025}, and $\sim 60\%$ of broad-\Halpha{} selected LRDs show significant Balmer absorption features \citep[e.g.,][]{Matthee2026}.
In general, the overall UV-to-optical spectrum of LRDs cannot be simply be explained by a combination of stellar populations, AGN and dust obscuration \citep[e.g.,][]{ma2025, degraaff2025, ji2025}.

Radiative transfer models that involve obscuration by a layer of dense, partly ionized gas ($n_{\rm H}\sim 10^8$--$10^{10}$~\unit{cm^{-3}}) with a high column density ($N_{\rm H}\sim10^{23}$--$10^{24}$~\unit{cm^{-2}}), have been successful in simultaneously explaining the strong Balmer breaks, Balmer absorption and strong Balmer line emission \citep[e.g.,][]{inayoshi2025, ji2025,naidu2025, degraaff2025, Torralba2026_gn9771,sneppen26} as well as the lack in X-rays \citep[e.g.,][]{kocevski2024,maiolino2025}. These models can also explain other features as a result of collisional and radiative transfer effects in a dense gas such as the strong [\ion{Fe}{ii}] \citep[e.g.,][]{Torralba2026_gn9771, deugenio2025_irony} or \ion{O}{i} emission lines \citep[e.g.,][]{tripodi_deep_2025, kokorev2025_glimpse_lrd}. A dense-gas envelope scenario also potentially explains the lack of short-term variability, by invoking super-Eddington sustained dense-gas flows \citep[e.g.,][]{secunda2025, Liu2026_atmosphere, madau_maiolino_2026}.
However, a coherent picture of the geometry, origin and composition of this gas, as well as the relation to the powering engine is under debate \citep[e.g.][]{InayoshiHo25,nandal2025,Chisholm26,madau_maiolino_2026}.

The LRDs are abundant at high redshift. With number densities of $\sim\,10^{-4}$~Mpc$^{-3}$ at $z\approx 4$--7 \citep[e.g.,][]{kocevski2023, maiolino2024a, greene2024, matthee2024a}, they represent a few percent of the galaxy population. They significantly outnumber other types of faint AGNs such as dust-obscured quasars as GNz7q \citep[e.g.,][]{fujimoto2022,Fei26} or faint X-ray detected sources at these redshifts \cite[e.g.,][]{Mahabal2005}.
Given this abundant population of sources with an unusual combination of spectral features at high-redshift unveiled by JWST, a natural question is to ask what evaded their identification in large spectroscopic surveys from the local Universe to cosmic noon (i.e., $z\lesssim 2$). At lower redshifts $2\lesssim z\lesssim4$, number density estimates of LRDs appear to drop sharply to $\Phi_{\rm LRD}\sim{\rm few}\times 10^{-6}\;{\rm Mpc^{-3}}$ and are therefore outnumbered by the X-ray AGNs $\Phi_{\rm AGN}\sim 10^{-4}\; {\rm Mpc^{-3}}$ \citep[e.g.,][vs. \citealt{ueda_x_ray_agn, Pouliasis_agn}]{loiacono2025,ma2025_counting}.
At even lower redshift, \citet{lin2026} performed a search for V-shaped broad line emitters in the Sloan Digital Sky Survey (SDSS) catalog, identifying three broad line sources as LRDs by their similarity to high-redshift counterparts (see also \citealt{ji2025_LoL}). Similarly to LRDs, these are X-ray faint \citep{Simmonds16}. Remarkably, two of these LRDs were already identified by \cite{Izotov07} during a search for metal-poor AGN, who discussed the impact of dense nuclear gas on the observed line-shape and line-ratios and their impact on the mass and spectrum of the ionizing source in \citep{Izotov08} akin to studies of high-redshift LRDs \citep[e.g.,][]{Matthee2026}. Despite challenges in the spectroscopic SDSS selection function, \cite{lin2026} estimated number densities of $\phi_{\rm LRD}[z<0.5]\gtrsim 5\times 10^{-10}\; {\rm Mpc^{-3}}$, further indicating a drop towards lower number densities \citep[see e.g.][]{inayoshi2025_decline}.

Besides studies of number densities that offer valuable constraints on the nature of LRDs \citep[e.g.,][]{Ma25noBright}, the proximity of low-redshift sources allows sensitive follow-up across the electromagnetic spectrum. For example, \cite{lin2026} were able to obtain high signal-to-noise follow-up spectroscopy, enabling the discovery of exceptionally strong metal absorption lines (Ca\,T, Na\,D, K\,{\sc i}) previously unknown in high-redshift LRDs.
Low-redshift samples also enable variability studies, over long time scales with ground-based facilities. 
At high-redshift, although observed at sparse cadence, LRDs do not show variability on $\lesssim1$ yr rest-frame time-scales \citep{Kokubo25,Tee25,liu2026}, disfavouring BLR scenarios with sub-Eddington accretion \citep{liu2026}.
\citet{burke2026} retrieved the light curves of three $z\sim 0.1$ LRDs in \citet{lin2025} from the Zwicky Transient Facility (ZTF) and Wide-field Infrared Survey Explorer (WISE) and have found that all three sources show weak $\sigma_0<0.03$ mag variability with hundreds of observations over rest-frame timescales of 10-20 years. In turn, tentative variability over long timescale was detected in a quadruply-lensed LRD with time delays between images being $\sim 130$ years showing photometric variability of 0.7 mag \citep{zhang2025}. %

Such examples highlight the advantages of low-redshift LRD studies, namely that ground-based (or other space-based) observatories can be utilized for both dense-cadence and multi-wavelength photometric monitoring (e.g., ZTF, LSST, WISE, HST, Chandra, etc) and high-resolution spectroscopy (VLT, etc), as well as significantly extending high-redshift constraints due to the lesser impact of redshift time dilation and data with baselines back a decade. 

We perform a systematic search in the DESI spectroscopic catalog \citep{DESI_DR1} to find additional local $z<0.5$ LRDs. Given its wide coverage of the sky (14 million galaxies in $A_{\rm DESI}\sim 14,000\; {\rm deg^2}$), the DESI catalog is an ideal dataset to find rare local LRDs. Unlike earlier searches at low-redshift \citep{ji2025_LoL,lin2025,Ding2026}, we do not necessarily perform a V-shaped selection. Instead, we select on broad emission-lines and then perform a template matching approach. The templates that we use encompass the diversity in SEDs of high-redshift LRDs that have been associated with a range in surrounding \ion{H}{i} gas column densities \citep[e.g.,][]{sneppen26,Matthee2026}. We note that such diversity is particularly prominent in the rest-frame UV to optical range (such as the strength of the Balmer break), and smaller in the rest-frame optical range where we template-match with the DESI spectra.

In Section~\ref{sec:Methods}, we describe how we performed our systematic search. In Section~\ref{sec:Results} we discuss the the eight LRDs we find in terms of the absorbers in the H$\alpha$ profiles, \ion{He}{i} $\lambda$5876 and \ion{He}{i} $\lambda$7067 strength, Balmer decrements, SEDs from archival photometry, and compactness. Moreover, we introduce BAQ 1, an intruiging quasar showing LRD-like Balmer absorption features. In Section~\ref{sec:variability}, we analyze the light curves of the LRDs and BAQ 1 and compare their variability properties. Finally, in Section~\ref{sec:discussion}, we estimate a number density of LRDs at low-redshift and discuss the implications of the fact that we do not find any luminous LRDs. We also compare our search method for local LRDs with others in the literature. Finally, in Section~\ref{sec:summary} we summarize our findings.

Throughout this paper we adopt a $\Lambda$CDM cosmology with $\Omega_m=0.3$, $\Omega_\Lambda=0.7$, and $H_0=70~\unit{km.s^{-1}.Mpc^{-1}}$. The magnitudes are given in the AB system \citep{oke1983}.

\section{Systematic Search for LRDs in DESI DR1}
\label{sec:Methods}

\begin{table}
\caption{Description of reduction of the DESI catalog via various selection criteria in our systematic search.}
\centering
\begin{tabular}{c|c}
\toprule
Selection Criterion & Remaining Sources \\
\midrule
\textit{Redshift cut} & \\
$z<0.45$ & 7,556,366\\ \hline 
\textit{Standard DESI broad line \& quality cut} & \\
\texttt{agn\_maskbit 12==True} & 72,689  \\
\texttt{halpha\_flux}$>5\times 10^{-17}\ {\rm erg\,s^{-1}\,cm^{-2}}$ & 66,893  \\
\texttt{hbeta\_flux}$>1\times 10^{-17}\ {\rm erg\,s^{-1}\,cm^{-2}}$& 65,360  \\ \hline
\textit{No strong lines with low-critical density} & \\
$\texttt{sii\_6731\_flux}/\texttt{halpha\_flux}<0.1$ & 13,262  \\
$\texttt{nii\_6584\_flux}/\texttt{halpha\_flux}<0.1$ & \textbf{3,078}  \\ \hline
\textit{LRD template selection} & \\
$\chi^2_{r,i}<100,\; i\in\{1,2,3,4\}$ & 381  \\
H$\alpha$ EW $>300 \AA$ & 210  \\
H$\alpha$ FWHM $>500$ ~\unit{km.s^{-1}} & 38  \\ \hline
\textit{Removing Quasars} & \\
{[\ion{Ne}{v}]} $\lambda3427\; {\rm EW}<3 \rm \AA$ & 19  \\
\hline
\textit{Final candidates} & \\
Ambiguous & 11 \\
{\bf LRDs}  & {\bf 8} \\
\bottomrule
\end{tabular}
\tablefoot{We highlight the 3,078 sources (which we call our parent sample) in bold after the [\ion{N}{ii}] criterion because it is the step in which we choose to inspect more carefully via template matching to the four LRD stacks of \cite{Matthee2026} and Gaussian fitting to individual spectral lines (H$\alpha$, [\ion{Ne}{v}]). }
\label{tab:selection_criteria}
\end{table}

We start our search for local LRD candidates in the DESI catalog by using the NOIR Data Lab \citep{datalab1,datalab2, datalab3} jupyter notebooks\footnote{\url{https://datalab.noirlab.edu/data-explorer}} and choose a redshift cut of $z_{\rm spec}<0.45$ to ensure that the H$\alpha$ emission line would be covered by the DESI wavelength range. Then we utilized the \texttt{agngal} value-added catalog\footnote{\url{https://data.desi.lbl.gov/doc/releases/dr1/vac/agngal/}} (Juneau et al., in prep) to select sources with broad and prominent Balmer emission lines (\texttt{agn\_maskbit 12: broad\_line}, a flag for a pipeline Gaussian FWHM $\geq 1200$~\unit{km.s^{-1}} in H$\alpha$, H$\beta$, \ion{Mg}{ii} and/or \ion{C}{iv} line) in galaxies and quasars and filter out stars with no emission line features. We also require a minimum line flux $F(\Halpha) > 5 \times 10^{-17}$~\unit{erg.s^{-1}.cm^{-2}} and $F(\Hbeta) > 1 \times 10^{-17}$~\unit{erg.s^{-1}.cm^{-2}}. Next, we remove sources with strong [\ion{S}{ii}] $\lambda6731$ and [\ion{N}{ii}] $\lambda6584$ line-emission compared to H$\alpha$ (see Table~\ref{tab:selection_criteria}). This is motivated by the typical spectra of known low-redshift LRDs (e.g., Fig.~8 in \citealt{lin2026}) and high-redshift LRDs \citep{Matthee2026}; their weak relative fluxes are likely related to their relatively low critical densities. 

These criteria narrow our sample from the $\sim 14$ million galaxies and quasars in DESI down to 3078 sources, which we name our "parent sample". We further pruned our parent sample via template fitting. We normalize all the spectra by median flux in the rest-frame 5100-5500$\AA$ range, and compute the reduced chi-squared $\chi^2_r$ against the spectra of the four high-redshift LRD stacks sorted by their Balmer break strength $f_{\nu, 5500}/f_{\nu, 3600}$ taken from \cite{Matthee2026}, which are normalized the same way. These stacks are based on JWST/NIRSpec grating data with comparable resolution as the DESI data (as opposed to NIRSpec PRISM spectra). All calculations ($4\times 3078\; \chi^2_r$ operations) are done in the rest-frame of the sources and in the range of rest-frame wavelength 4800-7500$\AA$, which is approximately the range where both DESI spectra and JWST stacks have good data quality. To choose a threshold for $\chi^2_r$, we compute $\chi^2_r$ against the four stacks for the three well-established local LRDs of \cite{lin2026} "J1047+0739", "J1022+0841", and "J1025+1402". We find that the minimum $\chi^2_r$ value across the stacks is 26.1, 6.3, 40.8 for each LRD; thus we (somewhat arbitrarily) choose $\chi^2_r\leq 100$ (for at least one of the stacks) as our criterion for our LRD classification. By fitting two-component Gaussians to the H$\alpha$ line, we require H$\alpha$ EW $>300 \AA$ and H$\alpha$ FWHM $>500$~\unit{km.s^{-1}} (of the total profile). This narrowed down our sample to 38 sources.

\begin{figure*}
    \centering
    \includegraphics[width=\linewidth]{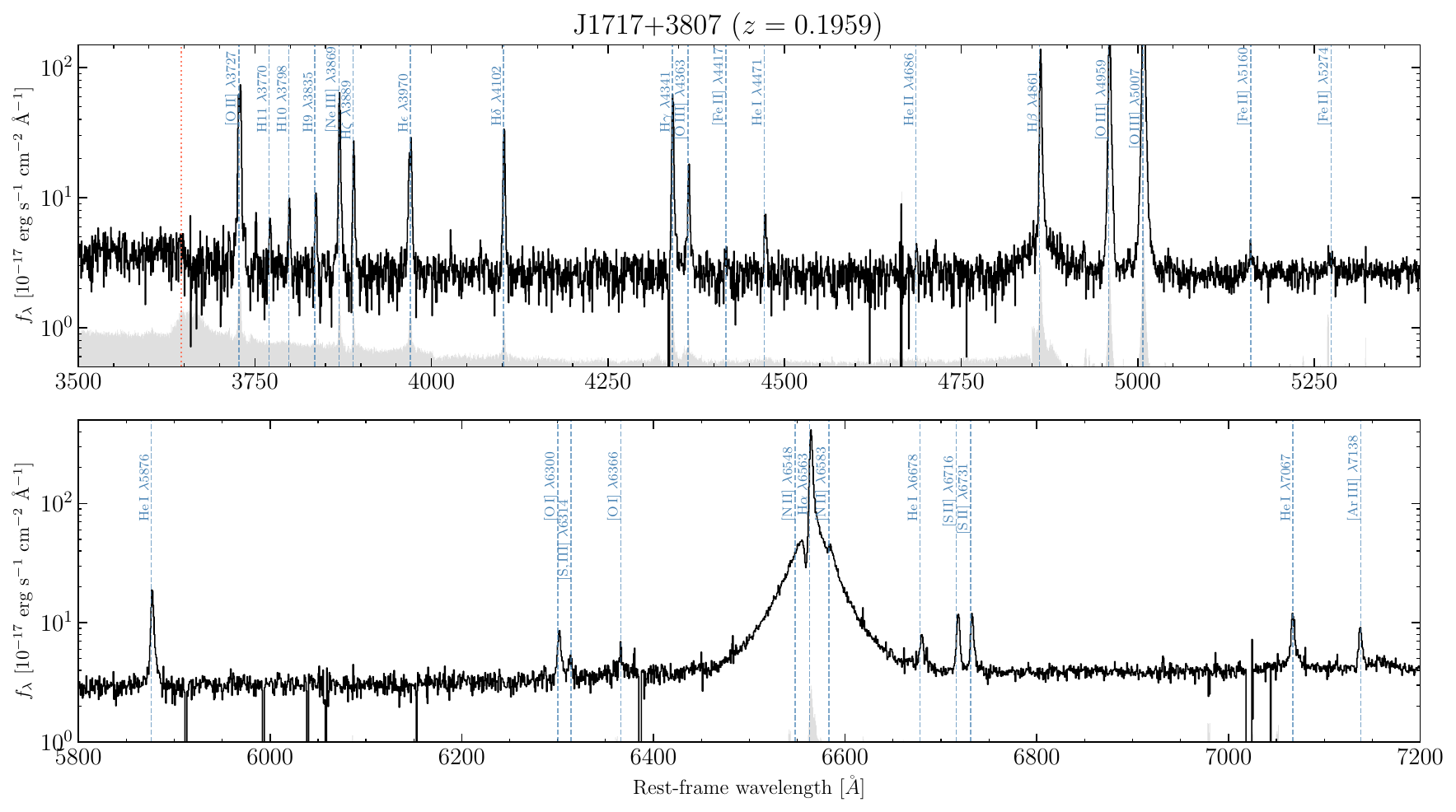}
    \caption{The DESI rest-frame spectrum of J1717+3807 (black), showing the rest-frame wavelength ranges 3500--5400~\AA{} (top) and 5800--7200~\AA{} (bottom). We mark the wavelengths of relevant spectral lines with blue dashed lines. As required by our search, the source shows weak [\ion{S}{ii}] and [\ion{N}{ii}], weak [\ion{Ne}{v}], strong \ion{He}{i} 5876, 7067, and a sharp absorption feature in the H$\alpha$ line. Intruigingly, we identify a Balmer jump (see red-dashed line) indicating nebular emission.}
    \label{fig:J1717_spectrum}
\end{figure*}

We discard sources with signatures of classical Type I broad-line AGN that are not seen in high-redshift LRDs from our sample. In particular, we remove objects with strong [\ion{Ne}{v}] $\lambda3427$ emission line ($\rm EW > 3 \AA$). This line has a high ionization potential (97~eV) and often correlates with strong X-ray emission of AGN origin \citep[e.g.,][]{Reiss2025, wang2026_AGN_NeV}, but it is not seen in LRDs \citep[as already discussed in][]{Izotov08}. We require our best candidates to present at least prominent \ion{He}{i} $\lambda5876$ and $\lambda7067$ lines and/or absorption features in the Balmer lines, typical signatures of high-$z$ LRDs \citep[e.g.,][]{Matthee2026}. Using these criteria, we identified 8 LRDs in the DESI dataset. We display the spectrum of the brightest source among the eight in Fig.~\ref{fig:J1717_spectrum} and mark relevant lines. We are unable to confirm or rule out a sample of 11 ambiguous sources as LRDs. Visual inspection of these sources indicates that this is primarily due to a low sensitivity that allows better $\chi^2$ values and/or prevents significant identification of strong [\ion{Ne}{v}] emission. Generally, we think that it is more likely that the majority of ambiguous sources are not LRDs with higher quality data.

Table~\ref{tab:selection_criteria} summarizes the remaining sources after each step in our systematic search. The source IDs and coordinates of the selected LRDs are listed in Table~\ref{tab:basic_properties}, whereas those of the ambiguous sources is in Appendix Table~\ref{tab:ambiguous1}. Fig.~\ref{fig:chi_sq_hist} shows their $\chi^2_r$ values of the LRD relative to the parent sample.

During the exploration phase that defined our search criteria, we also found a rare source which we dub "Balmer Absorption Quasar 1" (hereafter BAQ 1), with an \Halpha{} line profile remarkably similar to high-$z$ LRDs. However, this source has $\chi^2_r>100$ for all of the four templates, and was thus removed from our final sample. However, given that this source provides an opportunity to examine the differences between quasars and LRDs, we discuss its properties further in Section~\ref{sec:Results}.

We remark that the DESI pipeline classified all 8 LRD candidates as "GALAXY", even though they have a compact morphology. BAQ1 was indeed classified as a quasar. We provide hyperlinks to DESI interactive viewers in Table~\ref{tab:basic_properties} for these sources so that the reader can conveniently check basic information such as these classifications and the spectra, with relevant spectral lines marked, can also be found in Fig.~\ref{fig:J1717_spectrum}, Fig.~\ref{fig:J0129_J1343_full}, Fig.~\ref{fig:J1137_J1909_full}, Fig.~\ref{fig:J0829_J0716_full}, and Fig.~\ref{fig:J1502_full}.

\begin{figure}
    \centering
    \includegraphics[width=\linewidth]{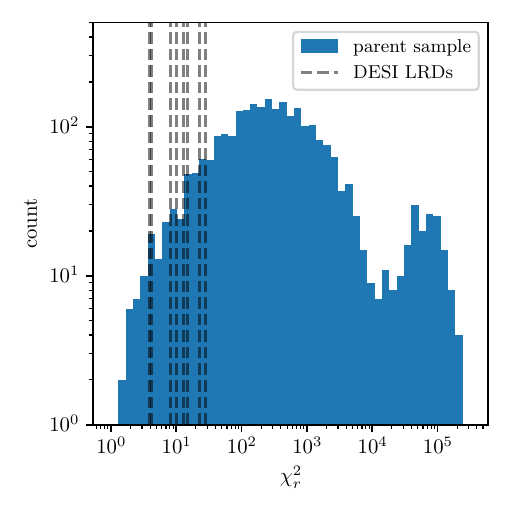}
    \caption{The reduced chi-squared distribution (minimum value across the four stacks) of the parent sample of N=3078 of our LRD search in DESI spectra (selected for broad emission lines, moderate H$\alpha$ and H$\beta$ flux, relatively weak [\ion{S}{ii}] and [\ion{N}{ii}] emission). We mark the locations of the eight LRD candidates we identified and the exact numerical values can be found in Table I. The second peak in the distribution at $\chi^2_r\sim10^5$ consists mostly of quenched galaxies.}
    \label{fig:chi_sq_hist}
\end{figure}

\section{Sample spectral properties}
\label{sec:Results}

\begin{table*}
\caption{Table of basic properties of the local LRDs of this work. }
\centering
\begin{tabular}{ccccccccc}
\toprule
Name & $z$ & $r$ (mag) & RA & DEC & Exp. time (s) & $\chi^2_{r}$ & $L_{\rm H\alpha}\, (10^{42}\; {\rm erg\,s^{-1}})$ & Break Strength \\
\midrule
\href{https://www.legacysurvey.org/viewer/desi-spectrum/dr1/targetid39633045061370969}{J1717+3807}& 0.1959 & 18.97 & 259.4239 & 38.1312 & 883 & 28.0 & $5.32 \pm 0.02$ & $0.97\pm0.03$\\
\href{https://www.legacysurvey.org/viewer/desi-spectrum/dr1/targetid39627941918802007}{J0129+0628} & 0.2467 & 21.22 & 22.3786 & 6.4787 & 1493 &14.7  & $1.55 \pm 0.01$ & $0.84\pm0.10$\\
\href{https://www.legacysurvey.org/viewer/desi-spectrum/dr1/targetid39633067916136674}{J1343+3934} & 0.2933 & 21.66 & 205.8242 & 39.5717 & 1441 & 12.8 & $1.08 \pm 0.02$ & $0.89\pm0.15$\\
\href{https://www.legacysurvey.org/viewer/desi-spectrum/dr1/targetid39633325291209396}{J1137+5520} & 0.4358 & 22.75 & 174.3931 & 55.3412 & 1632& 4.2 &  $1.57 \pm 0.10$ & $2.05\pm0.60$\\
\vspace{0.15cm}
\href{https://www.legacysurvey.org/viewer/desi-spectrum/dr1/targetid39633369390125151}{J1909+5831} & 0.4273 & 22.85 & 287.4756 & 58.5201 & 4024 &3.8 & $0.97 \pm 0.12$ & $1.68\pm 1.07$\\
\href{https://www.legacysurvey.org/viewer/desi-spectrum/dr1/targetid39628104418725302}{J0829+1312} & 0.3986 & 20.63 & 127.3390 & 13.2104 & 1182 &8.2 &  $5.01 \pm 0.06$ & $0.85\pm0.08$\\
\href{https://www.legacysurvey.org/viewer/desi-spectrum/dr1/targetid39633314180500467}{J0716+5433} & 0.2908 & 20.63 & 109.1489 & 54.5561 & 1456 &22.9 &  $2.61 \pm 0.02$ &$1.20\pm0.07$\\
\href{https://www.legacysurvey.org/viewer/desi-spectrum/dr1/targetid39627854949912902}{J1502+0250} & 0.2906 & 21.42 & 225.7181 & 2.8411 & 1307 &10.2 &  $1.21 \pm 0.02$ & $1.30\pm0.13$\\

\bottomrule
\end{tabular}
\tablefoot{
The columns correspond to the name of the source, redshift, $r$ magnitude, J2000 coordinates, DESI exposure time, reduced chi-squared relative to the best of the four high-redshift LRD stacks, H$\alpha$ luminosity, and the Balmer break strength. The break strength is defined as $f_{\nu, 5500}/f_{\nu, 3600}$. We placed LRDs with absorption features in their H$\alpha$ line (J1717+3807, J0129+0628, J1343+3934, J1137+5520, J1909+5831) first in order of $r$ band magnitude, then LRDs without absorption features (J0829+1312, J0716+5433, J1502+0250) in order of $r$ band magnitude. The first column includes clickable links to the DESI spectrum viewer.}
\label{tab:basic_properties}
\end{table*}

\begin{table*}
\caption{Properties of the local LRDs identified in this work (continued from Table~\ref{tab:basic_properties}) .}
\centering
\begin{tabular}{lccccccc}
\toprule
Name &  ${\rm FWHM}_{\rm exp, H\alpha}$ & H$\alpha$/H$\beta$ & H$\alpha$ EW & \ion{He}{i}$_{5876}$ EW & \ion{He}{i}$_{7067}$ EW & $\tau_{0, \rm abs}$ & $\Delta$v$_{\rm abs}$ \\
\midrule
\href{https://www.legacysurvey.org/viewer/desi-spectrum/dr1/targetid39633045061370969}{J1717+3807} & $1248 \pm 10$ & $8.9\pm1.6$ & $985\pm 22$ & $26.6\pm4.2$ & $8.3\pm0.4$ & $3.2 \pm 0.2$ & $-230 \pm 6$ \\
\href{https://www.legacysurvey.org/viewer/desi-spectrum/dr1/targetid39627941918802007}{J0129+0628} & $748 \pm 9$ & $11.0\pm 1.6$ & $994\pm 36$ & $20.6\pm1.3$ & $16.7\pm 0.9$ & $4.1 \pm 1.5$ & $-66 \pm 8$ \\
\href{https://www.legacysurvey.org/viewer/desi-spectrum/dr1/targetid39633067916136674}{J1343+3934} & $588 \pm 16$ & $5.3\pm1.2$ & $1088\pm 67$ & $54.2\pm7.2$ & $35.0\pm2.5$ & $5.2 \pm 0.6$ & $-99 \pm 8$ \\
\href{https://www.legacysurvey.org/viewer/desi-spectrum/dr1/targetid39633325291209396}{J1137+5520} & $891 \pm 39$ & $8.9\pm1.9$ & $760\pm69$ & $19.5\pm 1.8$ & N/A & $4.2 \pm 2.8$ & $-143 \pm 19$ \\
\vspace{0.15cm}
\href{https://www.legacysurvey.org/viewer/desi-spectrum/dr1/targetid39633369390125151}{J1909+5831} & $748 \pm 37$ & $9.6\pm4.6$ & $423\pm 57$ & $11.3\pm1.9$ & N/A & $3.8 \pm 5.2$ & $-75 \pm 24$ \\
\href{https://www.legacysurvey.org/viewer/desi-spectrum/dr1/targetid39628104418725302}{J0829+1312} & $944 \pm 11$ & $7.4\pm 0.8$ & $1082\pm 41$ & $25.1\pm1.0$ & N/A & N/A & N/A \\
\href{https://www.legacysurvey.org/viewer/desi-spectrum/dr1/targetid39633314180500467}{J0716+5433} & $1258 \pm 15$ & $7.7\pm1.2$ & $1206\pm46$ & $18.3\pm0.9$ & $10.5\pm 0.6$ & N/A & N/A \\
\href{https://www.legacysurvey.org/viewer/desi-spectrum/dr1/targetid39627854949912902}{J1502+0250} & $685 \pm 11$ & $13.2\pm3.1$ & $493\pm 13$ & $8.0\pm 0.9$ & $3.5\pm 0.5$ & N/A & N/A \\
\bottomrule
\end{tabular}
\tablefoot{The H$\alpha$ and HeI EWs are in {$\AA$}, the FWHM of the exponential component is in km s$^{-1}$, $\tau_{0, \rm abs}$ is the depth of the H$\alpha$ absorber and its velocity offset ($\Delta$v$_{\rm abs}$) is in km s$^{-1}$, where the offset is defined with respect to the center of the exponential component.}
\label{tab:halpha_properties}
\end{table*}

\subsection{H$\alpha$ lines and absorbers}\label{sec:ha_model}

\begin{figure*}
    \centering
    \includegraphics[width=\linewidth]{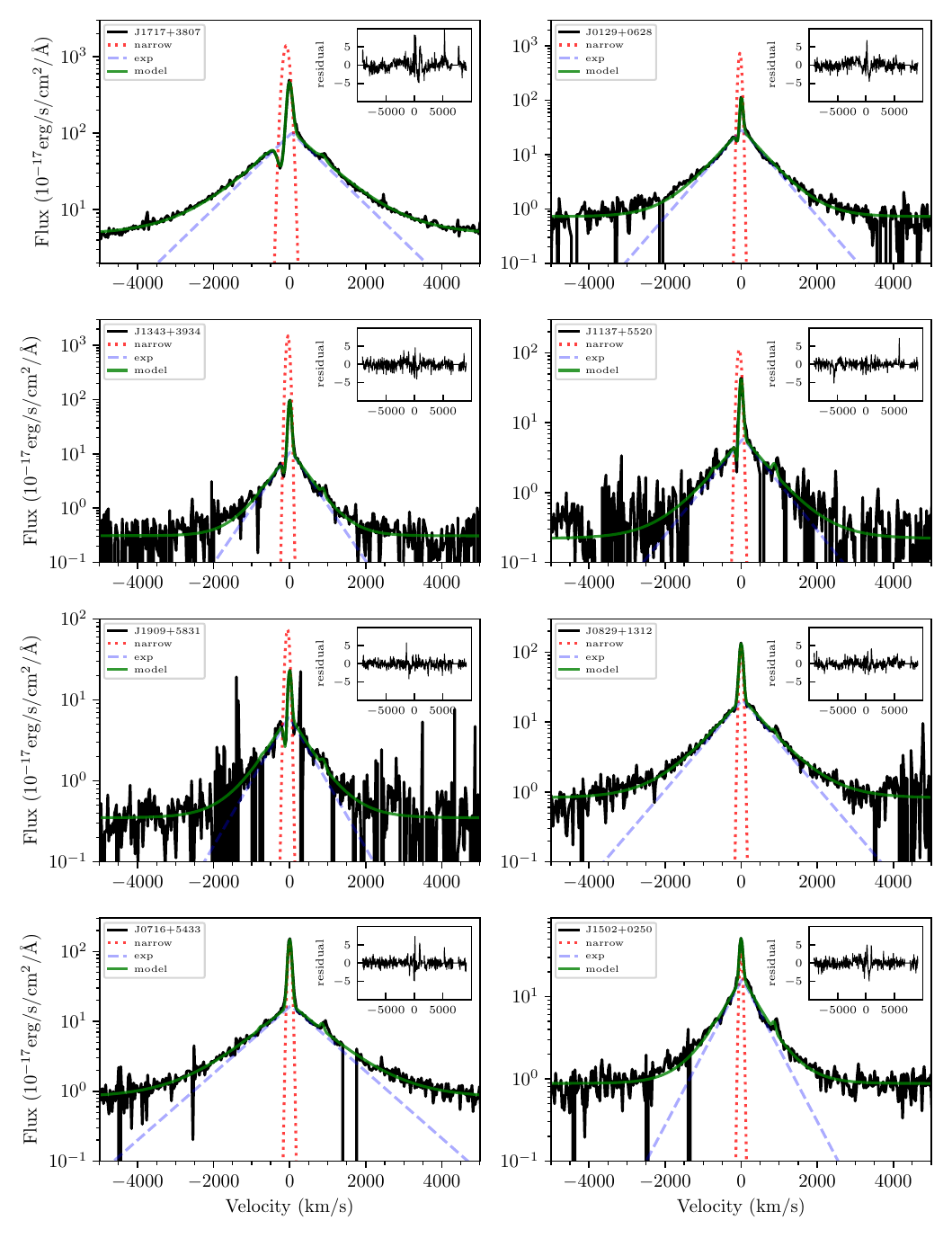}
    \caption{H$\alpha$ profiles of all local LRDs identified in this work, along with the best-fit model in green, the exponential component of the model in dashed-blue, and the narrow Gaussian component in dashed-red. The residuals in the insets show ${\rm (data-model)/error}$.}
    \label{fig:ha_all}
\end{figure*}

Here we investigate how the detailed H$\alpha$ line profiles of LRDs compare to  high-$z$ LRDs, that display a combination of broad wings ($\rm FWHM > 1000$~\unit{km.s^{-1}}) and a narrow/intermediate component near the line center, with 5 out of 8 sources presenting absorption features \citep[see e.g.,][]{matthee2024a, Matthee2026, Torralba2026_gn9771, deugenio2025, deugenio2025_irony, deugenio2025_absorber, ji2025}.
Following the methodology in \citet{Torralba2026_gn9771}, we fit the H$\alpha$ profiles of the local LRDs with a model composed of a narrow Gaussian $N(\lambda)$, broad symmetric exponential $E(\lambda)$, two Gaussians representing the [\ion{N}{ii}] doublet (with a fixed theoretical ratio of 2.96), and an absorber parametrized as ${\rm e}^{-\tau_{\rm abs}}$, where $\tau_{\rm abs}(\lambda)$ follows a Gaussian distribution. These components are then convolved with a line-spread function represented by a 1D Gaussian with width (${\rm FWHM}=\lambda_{H\alpha}/R_{\rm DESI}$) using a representative resolution of DESI in the infrared ($R=4100$--5000 for $\lambda=6560$--$9800$\footnote{\url{https://www.desi.lbl.gov/spectrograph/}}; we choose the midpoint $R_{\rm DESI}\approx4550$). The continuum around the emission line is modeled by a constant for simplicity and we mask the region around the [\ion{S}{ii}] doublet and \ion{He}{i}$_{6678}$. In summary, we fit the profiles to the following model:
\begin{equation}\label{eq:Ha_model}
    f_\lambda(\lambda) = \left\{\left[ N(\lambda) + E(\lambda) \right] \cdot {\rm e}^{-\tau_{\rm abs}(\lambda)} + {\rm cont.} + [\ion{N}{ii}]\right\} * {\rm LSF} \,.
\end{equation}

Calculations were done using the least-squares method of the non-linear curve-fitting Python package \texttt{lmfit} \citep{lmfit}. The H$\alpha$ profiles and the best-fit model are shown in Fig.~\ref{fig:ha_all}. Overall, we find good agreement between our model and the data, with $\chi^2_{r, \rm model}=1$--6. The broad wings are well described by a single symmetric exponential \citep[see also e.g.][]{rusakov2025}, and these are topped by a P Cygni profile \citep[e.g.,][]{Matthee2026} that is characteristic of an outflowing partially ionised medium. Using this model we measured the integrated H$\alpha$ flux, the FWHM from the exponential component, the optical depth $\tau_0$ of the absorption (i.e., the peak of the Gaussian representing the absorber $\tau_{\rm absorber}(\lambda)$), and the velocity offset of the center of the absorber relative to the center of the broad exponential, which we show in Table~\ref{tab:halpha_properties}. For our sources, we find FWHM values of the exponential broad components of $\sim 600$--$1200$~\unit{km.s^{-1}}.

In 5 out of our 8 LRDs, we see significant absorption features, with the fiducial model in Eq.~\ref{eq:Ha_model} statistically preferred ($\Delta {\rm BIC}\gtrsim 80$) over a model not including the absorption factor. Assuming that the velocity of the absorber is Gaussian-distributed, our models detect outflows (blueshifted absorption) ranging from $63$--$286$~\unit{km.s^{-1}} with line-center optical depths of $\tau_0=1$--$5$. While the absorption features for J1717+3807, J0129+0628, J1343+3934 are clear, the overall S/N of the spectra of J1137+5520 and J1909+5831 are lower because they are near the detection limit of DESI with $r$-band magnitudes of $\sim 22.8$ with the faintest H$\alpha$ fluxes among our sample (see Table~\ref{tab:basic_properties}). Although the absorption features are not correlated with the location of noise spikes, a more precise characterization of the absorption will likely require deeper follow-up spectroscopy with higher spectral resolution. In particular, we speculate that the H$\alpha$ profile of J1909+5831 could also show redshifted absorption, thus featuring simultaneous redshifted and blueshifted absorption similar to the LRD in \cite{deugenio2025} or a single deep absorption trough \citep[e.g.,][]{torralba2026_bh_star}. However, given that the redshifted absorption is next to a noise spike, obtaining follow-up optical spectroscopy for J1909+5831 is necessary to make more robust conclusions.

\begin{figure}
    \centering
    \includegraphics[width=\linewidth]{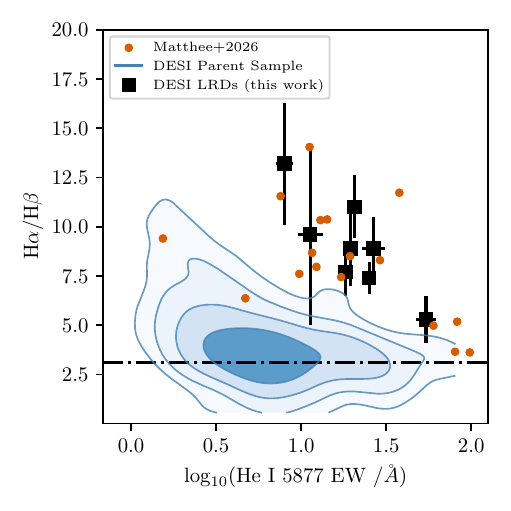}
    \caption{Scatter plot of \ion{He}{i} $\lambda5876$ EW vs H$\alpha$/H$\beta$ ratios for our parent sample in DESI (blue contours) compared to high redshift LRDs of \cite{Matthee2026} (red points) and the local LRDs of this work (black points). DESI contours show a Gaussian KDE, with contour levels corresponding to $e^{-1}$, $e^{-2}$, $e^{-3}$, and $e^{-4}$ times the peak density. We mark the theoretical Case B recombination value of 3.1 with a horizontal dashed line along with the positions of local LRDs marked with black points. We observe a clustering of non-LRD sources around this value and an anti-correlation between the H$\alpha$/H$\beta$ ratios and \ion{He}{i} $\lambda5876$ EWs for LRDs. }
    \label{fig:HaHbratio_and_HeI}
\end{figure}

\subsection{HeI strength and H$\alpha$/H$\beta$ ratios}\label{sec:hahb_hei}

Our LRDs have strong \ion{He}{i} emission, analoguous to high-redshift counterparts, with EWs of \ion{He}{i} $\lambda5876$ typically being $\sim 20\AA$ and reaching up to $52.5 \AA$ (and $33.6\AA$ for \ion{He}{i} $\lambda7067$) for J1343+3934. The relative intensity of the triplet He I lines (such as $\lambda5876$ and $\lambda7067$) is sensitive to radiative-transfer processes in a very dense gas \citep[e.g.,][]{benjamin2002, berg2025}. The triplet ground level $2^3$S of \ion{He}{i} is depopulated at high densities \citep[e.g.,][]{mathis1957, benjamin2002}, boosting the intensities of such lines via collisional excitation to higher triplet levels followed by radiative cascades \citep[e.g.,][]{berg2025, Torralba2026_gn9771}. By fitting Gaussians to the \ion{He}{i} $\lambda$5876 and \ion{He}{i} $\lambda$7067 emission lines of our parent sample, we calculate and show the equivalent widths (EWs) in Table~\ref{tab:basic_properties}. For $z> 0.35,$ the \ion{He}{i} $\lambda7067$ line gets redshifted into wavelengths that are not covered by DESI.

Another emission-line ratio that is particularly noteworthy is the H$\alpha$/H$\beta$ ratio (the Balmer decrement), which is sensitive to the dust attenuation \citep[e.g.,][]{cardelli1989}, but also to collisional excitation \citep[e.g.,][]{raga2015, Torralba2026_gn9771, Berg26} and resonant scattering effects in dense media \citep{Chang25}.  By fitting double Gaussians to the H$\alpha$ and H$\beta$ lines in our parent sample, we also measure the total H$\alpha$/H$\beta$ flux ratios. In the standard case B scenario for typical BLR conditions, the H$\alpha$/H$\beta$ ratio is predicted to be $3.1,$ whereas large H$\alpha$/H$\beta$ ratios H$\alpha$/H$\beta \sim9$ are commonly found for high redshift LRDs and explained with radiative-transfer or collisional effects in dense gas \citep[e.g.,][]{nikopoulos2025,Chang25, degraaff25BH*,Matthee2026}, or dust attenuation \citep[e.g.,][]{Brooks25}. 

As shown in Figure~\ref{fig:HaHbratio_and_HeI}, we find consistent results with the high-redshift LRDs, with most sources showing large ($\gtrsim 7$) H$\alpha$/H$\beta$ ratios. Meanwhile, in the parent sample of broad-line AGN with weak [\ion{S}{ii}] and [\ion{N}{ii}], we find a clustering of points near H$\alpha$/H$\beta \approx3.1$. Moreover, the LRDs stand out significantly in terms of their \ion{He}{i} EW, similar to the high-redshift sources \citep{Matthee2026}. Both the low-redshift and high-redshift LRDs display an anti-correlation between the Balmer deecrement and the EW of the HeI line.

\subsection{Multi-wavelength photometry}
\label{sec:multiwavelength_phot}

\begin{figure*}
    \centering
    \includegraphics[width=\linewidth]{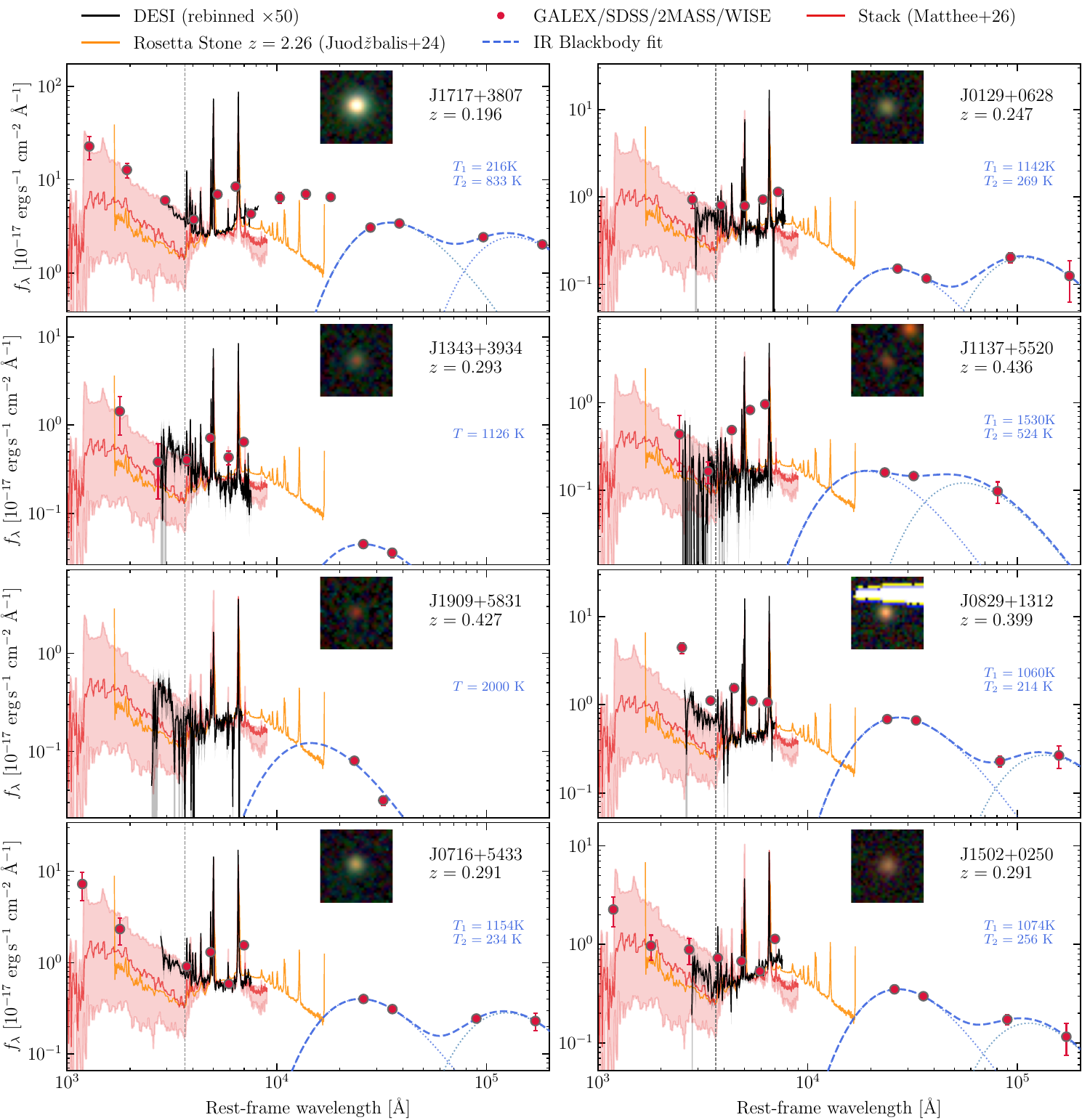}
    \caption{Spectral energy distributions of our sample. We show the DESI spectrum with a black line, resampled to a $\times50$ coarser binning for visual clarity. We also show archival photometry from GALEX, SDSS, 2MASS and WISE with red dots. We also compare with the stack of the full JWST sample in~\citet{Matthee2026} (red line and shaded region) and the JWST prism spectrum of the \textit{Rosetta Stone} at $z=2.26$ \citep{juodzbalis2024} that covers the rest-frame NIR. We also show our best-fit blackbodies to the mid-IR photometric points (rest wavelength $>2$~\unit{\mu m}; see Sect.~\ref{sec:multiwavelength_phot}). We also show inset RGB (DESI legacy $g$, $r$, and $z$ bands) $7.86\arcsec\times7.86\arcsec$ stamps of each object, showcasing their compact morphology.}
    \label{fig:seds8}
\end{figure*}

As mentioned in Sect.~\ref{sec:introduction}, local LRDs open opportunities for ground-based observatories to contribute to the study of LRDs. We query the VizieR photometry viewer\footnote{\url{https://vizier.cds.unistra.fr/vizier/sed/old/}} for the local LRDs to retrieve archival data. With the exception of J1909+5831, which is the faintest of our sources, we find photometry covering rest-UV \citep[GALEX;][]{bianchi2017}, optical \citep[SDSS;][]{abdurrouf2022}, and infrared \citep[2MASS and WISE;][]{Skrutskie2006, Wright2010}. We find no data or constraints in X-rays; in particular, to our knowledge, the 8 sources lie in areas of the sky where Chandra has not observed. 

We convert the flux densities to the rest-frame using the redshift measured by DESI and plot them along with the DESI spectra in Fig.~\ref{fig:seds8}. To check if a Balmer break-like inflection point occurs, we mark the Balmer break wavelength and also plot the stack of \cite{Matthee2026} for reference. We find that J1717+3807 and J1502+0250 show the clearest inflection points, both in the photometry and in the smoothed spectra. With more uncertainty, J0829+1312 also shows similar trends. For sources like J1343+3934, J1137+5520, and J0129+0628, the optical photometry frequently overlaps with strong emission lines, making analysis of the continuum difficult.

In all sources, we find weak IR emission, consistent with the high-redshift LRDs \citep[e.g.,][cf. \citealt{delvecchio_active_2025, barro2024}]{Setton2025}. In order to roughly quantify this, we fitted two blackbodies to the photometric points at $\lambda_{\rm rest} > 2$~\unit{\mu m} (with the exception of J1909+5831, for which we fit a single blackbody since only two data points are available). As shown in Fig.~\ref{fig:seds8}, some contribution to the mid-IR continuum could arise from a ``warm dust'' emission component with $T\sim 200$--1100~\unit{K}.

We also produce RGB images of our sources using the DESI legacy $g$, $r$, $z$ bands and show $7.86\arcsec\times7.86\arcsec$ cutouts in Fig.~\ref{fig:seds8}. We note that despite never explicitly requiring a compactness requirement in our spectroscopic search, the final eight sources appear unresolved, implying angular sizes of $\lesssim 1\arcsec.$ For the closest source J1717+3807 at $z=0.1959$, this implies a comoving physical size of $\lesssim 3.8\; {\rm kpc}$ and for the farthest source J1137+5520 at $z\sim 0.4358$, this implies a size of $\lesssim 8.1 \;{\rm kpc}.$ Given that the sources are unresolved, the true sizes could be much smaller; thus, we only interpret these numbers as an upper limit.

\subsection{Stacks} \label{sec:stacks}

\begin{figure*}
    \centering
    \includegraphics[width=\linewidth]{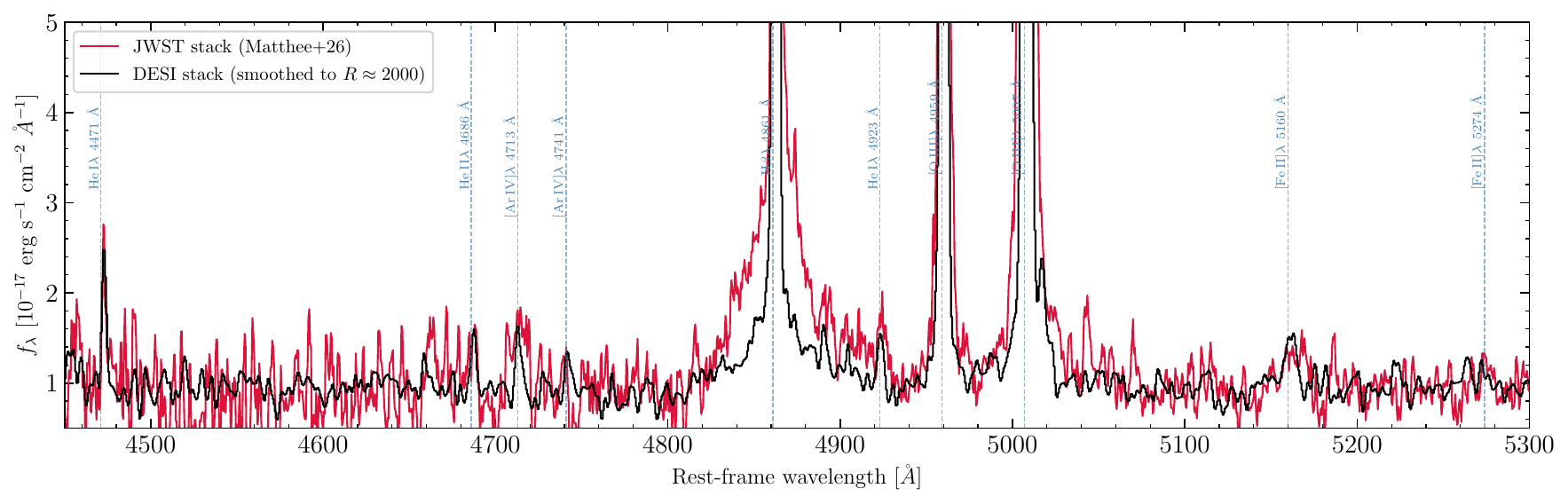}
    \includegraphics[width=\linewidth]{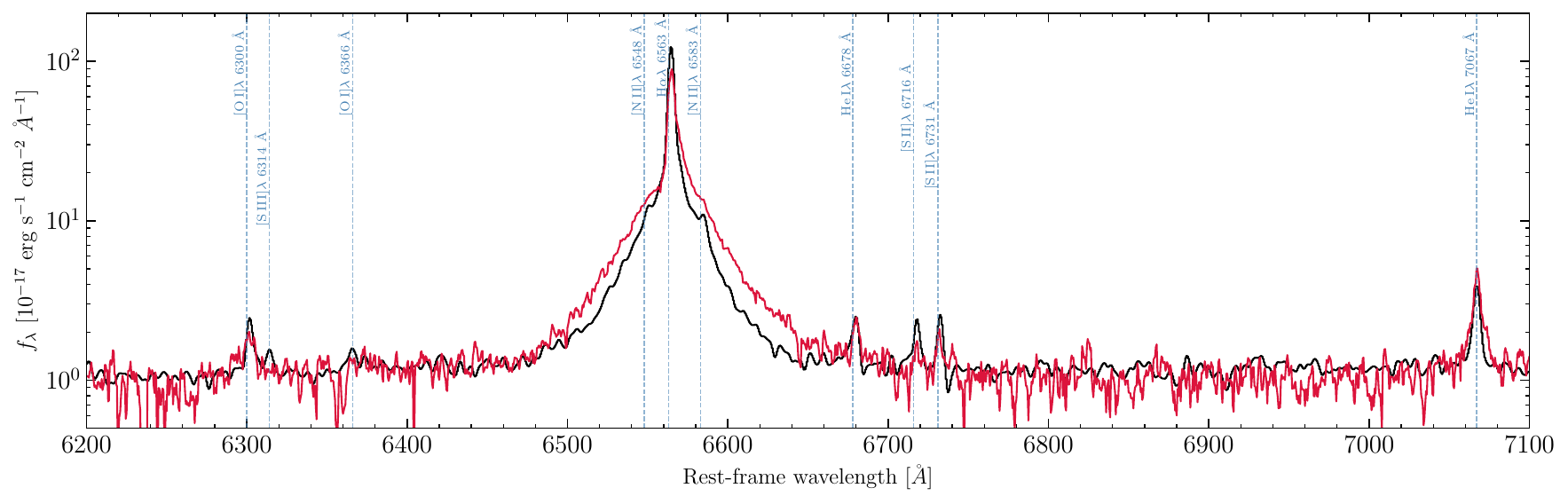}
    \caption{Stack of our 8 selected LRDs (black) compared to the JWST/NIRSpec grating stacks in \citet{Matthee2026} (red). We smoothed the DESI stack ($R_{\rm DESI}\approx 4550$) with a Gaussian kernel to match the NIRSpec stack (assuming $R_{\rm grating}\approx 2000$). The DESI stack shows similar narrow line features to the JWST counterpart; including [\ion{Fe}{ii}], \ion{He}{i}, [\ion{Ar}{iv}], and [\ion{S}{iii}]. The main difference is the relative intensity of the Balmer lines \Halpha{} and \Hbeta{}, this is discussed in Sect.~\ref{sec:stacks}.}
    \label{fig:stacks_full}
\end{figure*}

We stacked our eight selected LRDs to examine the general properties of the sample. In Fig.~\ref{fig:stacks_full} we show the median DESI stacks, compared to the stack of the full JWST/NIRSpec sample in \citet{Matthee2026}. For the comparison, we have degraded the DESI stack ($R_{\rm DESI}\approx 4550$) to the approximate resolution of the NIRSpec grating stack ($R_{\rm grating}\approx 2000$). Also, both stacks are normalized to their median flux in the $5100-5300 \AA$ wavelength range. The DESI stack qualitatively shows very similar narrow-line features to the JWST stack (in particular [\ion{Fe}{ii}], \ion{He}{i}, [\ion{Ar}{iv}], and [\ion{S}{iii}]). In turn, the main difference is the relative strength of the broad Balmer lines. 

We fit Gaussians to the \ion{He}{ii} $\lambda4868$ and [\ion{Ne}{v}] $\lambda3426$, which we imposed to be weak by selection (see Sect.~\ref{sec:Methods}). Both lines have relatively high ionization potentials (IP 54.4~eV and 97.1~eV, respectively), and ensure that our selected sources have intrinsically soft ionization spectrum, contrary to what is expected for broad like Type I AGN. For our stack, we measure rest-frame equivalent widths of $\rm EW(\ion{He}{ii}) = 2.3 \pm 0.4$~\AA{}, $\rm EW([\ion{Ne}{v}]) = 0.2 \pm 0.3$~\AA{} (3$\sigma$ upper limit $0.9$~\AA{}). We also fit the \Halpha{} and \Hbeta{} in our median stack with the same model we used in Sect.~\ref{sec:ha_model}. We obtain for our stacked sample $\rm EW(\Halpha) = 910 \pm 30$~\AA{}, $\rm EW(\Hbeta) = 170 \pm 9$~\AA{}, and $\rm EW([\ion{N}{ii}]) = 13 \pm 2$~\AA{} (doublet). For comparison, the median H$\alpha$ EW in the high-redshift stack is 1500 {\AA}, but we note that the \Halpha{} EW varies from source to source among LRDs \citep[300--2000~\AA{}; e.g.,][]{Hviding2025, degraaff25BH*}. .
The Balmer decrement of the stack is $\Halpha{}/\Hbeta{} = 7.3 \pm 0.7$ (see Sect.~\ref{sec:hahb_hei} for a detailed discussion).

Following \citet{Wang2025_ion}, we measure optical line ratios $\log_{10}(\ion{He}{ii}/\Hbeta) = -1.84 \pm 0.08$ and $\log_{10}([\ion{N}{ii}]/\Halpha) = -1.99 \pm 0.08$. Both ratios are much lower than typical values observed in SDSS Type I AGN \citep[e.g.,][]{Bar2017}, securing our stacked sample within the non-AGN regime \citep[e.g.,][]{Kouroumpatzakis2025}. Moreover, as discussed by \citet{Wang2025_ion} that the extremely high \Halpha{} equivalent widths measured in most LRDs (including our sample) is energetically incompatible with typical QSO ionization spectra. We can thus conclude that the emission-lines in our sample are not compatible with the ionization conditions of a typical BLR, and these are rather indicative of a hot ionising spectrum with a cutoff above $\sim60$ eV.

\subsection{Contaminants: Balmer absorption quasars}
\label{sec:balmer_absorption_qsos}

During the visual inspection of our parent sample of 3,078 sources (see Sect.~\ref{sec:Methods}), we identified 7 objects that show broad Balmer lines and Balmer absorption (see Fig.~\ref{fig:baq_has}) that are visually similar (to varying degrees) to the ones found in high-redshift LRDs \citep[e.g.,][]{Matthee2026}. Although the sources do not pass the $\chi^2_r<100$ cut, given the striking similarities of the H$\alpha$ profile, we discuss the properties of one of these sources that most resembles an LRD. 

\begin{figure}
    \centering
    \includegraphics[width=\linewidth]{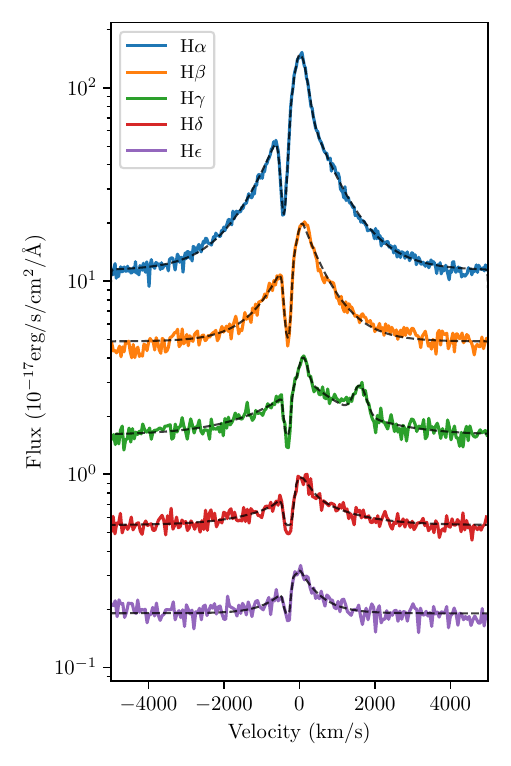}
    \caption{The Balmer lines (H$\alpha$--H$\epsilon$) of the LRD-contaminant BAQ1 that shows strong blueshifted absorption features at $\sim 350$~\unit{km.s^{-1}}. We overlay the spectra with our best-fit model (see model details in Section~\ref{sec:Results}) using dashed gray curves. In the case of H$\gamma$ we simultaneously fit the [\ion{O}{iii}]$\lambda 4363$ line. Spectra are offset vertically for clarity. }
    \label{fig:baq1_balmer}
\end{figure}

BAQ1\footnote{DESI interactive viewer for \href{https://www.legacysurvey.org/viewer/desi-spectrum/dr1/targetid39628324502246418}{BAQ1}.} (RA, Dec = (236.2971, +22.6489), $z=0.2187$, magnitude $r=18.47$) is an intriguing source that we discovered during our visual inspection. It caught our attention because it shows narrow, yet strong absorption features blueshifted by few hundred \unit{km.s^{-1}} relative to the broad component in its Balmer lines. The absorption goes below the continuum in the higher order lines. Using the same model as in Eq.~\ref{eq:Ha_model} (but accounting for [\ion{O}{iii}] $\lambda$4364 near the H$\gamma$ line with an additional single Gaussian), we quantify and show the characteristics of the absorber of this source in Table~\ref{tab:BAQ1_absorber}. We also show the Balmer lines \Halpha{}--H$\epsilon$ in Fig~\ref{fig:baq1_balmer}, along with the best-fit model in a gray-dashed line. The center of the absorption occurs roughly in the same location in velocity space along the Balmer series, indicating a common physical origin. We also note that the absorption optical depth is larger for H$\beta$ compared to H$\alpha$, which is unusual for typical galaxies, but has been reported in high redshift LRDs \citep[e.g.,][]{deugenio2025}. The central absorption velocities of $\approx-350$~\unit{km.s^{-1}} are more blue-shifted than typically seen in LRDs, suggesting faster moving gas. Moreover, the source shows negligible [\ion{N}{ii}] and [\ion{S}{ii}] compared to H$\alpha$ and shows prominent \ion{He}{i} emission. However, some features of BAQ1 are at odds with most observations of high-$z$ LRDs and are in turn typical signatures of the high-ionization photons and hard ionization spectra of classical broad line AGN. For example, we measure a [\ion{Ne}{v}] EW of $1.4\ \AA$ and we observe strong and broad \ion{He}{ii} emission. Notably, also the Balmer decrement is not as steep as in LRDs (${\rm H\alpha/H\beta}=3.7\pm0.3$).

\section{Variability analysis}
\label{sec:variability}

\begin{figure}
    \centering
    \includegraphics[width=\linewidth]{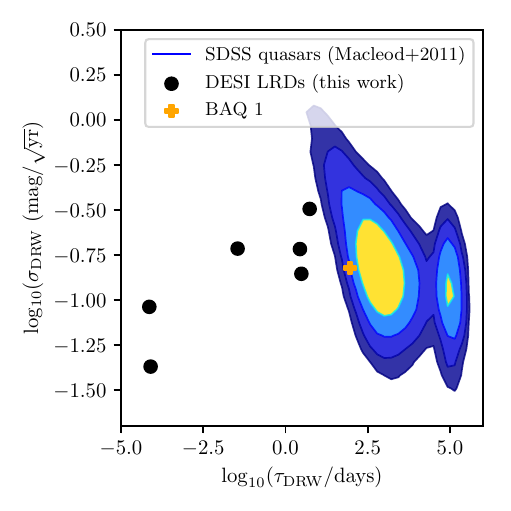}
    \caption{Measured Damped Random Walk (DRW) parameters (amplitude and correlation timescale) for the r-band light curves of 9258 SDSS quasars \cite{macleod_2012}, DESI LRDs from this work, and BAQ 1 from this work. The contours are estimated and plotted identically to Fig.~\ref{fig:HaHbratio_and_HeI} using Gaussian KDE.}
    \label{fig:DRW_comparison}
\end{figure}

\begin{figure*}
    \centering
    \includegraphics[width=\linewidth]{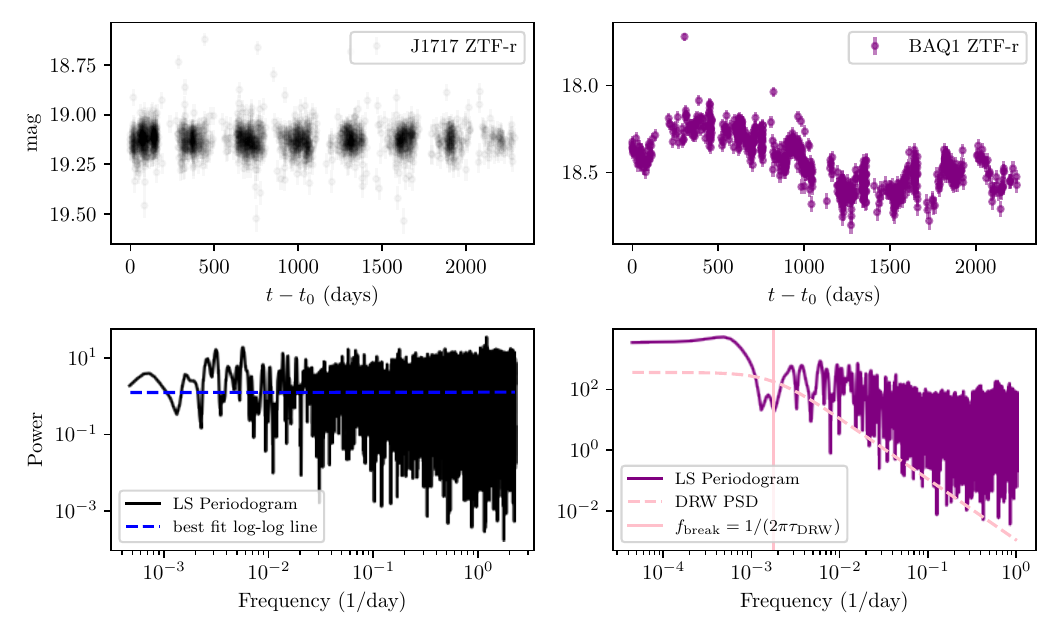}
    \caption{ZTF r-band rest-frame light curves of J1717+3801 and BAQ1 (top panels) along with their LS Periodograms (bottom panels) to estimate their PSDs. The data points of J1717+3801 are shown with high transparency to emphasize where they are concentrated the most. The PSD of J1717 is flat ($\propto f^0$), indicative of uncorrelated white noise, whereas the PSD of BAQ1 is bent ($\propto f^{-2}$), indicative of correlated red noise, with a flattening occuring roughly at the theoretical value of $f_{\rm break}=1/(2\pi \tau_{\rm DRW}).$ We find that the light curve of J1717+3801 shows signs of weak variability ($\sim 0.04$ mag) and uncorrelated white noise, whereas BAQ1 shows typical quasar variability of $\sigma_{\rm DRW}\sim 0.15$ mag.}
    \label{fig:lc_analysis}
\end{figure*}

 Seven out of the eight sources (the exception being J1137+5520) have light curves in ZTF\footnote{\url{https://irsa.ipac.caltech.edu/cgi-bin/Gator/nph-dd}} spanning up to $\sim2500$ days, or 7 years in the rest-frame. We choose J1717+3807 as a representative example for analysis as it is sufficiently bright to be detected in both CRTS\footnote{\url{http://nunuku.caltech.edu/cgi-bin/getcssconedb_release_img.cgi}} \citep{CRTS2009} and ZTF \citep{Masci_2019}, establishing a baseline of $\sim 17$ years in the rest-frame, and most of its data points are within the typical limiting magnitude of ZTF ($20.5$ mag). As can be seen in the ZTF light curve of this source in the left panel of Fig.~\ref{fig:lc_analysis}, J1717+3807 does not appear to vary. Subtracting out noise, we estimate the intrinsic variability of this source as $\sigma^2_{\rm int}={\rm var(y)}-\langle\sigma^2_{\rm err}\rangle$, where ${\rm var(y)}$ is the variance of the light curve $y$ in magnitude units and $\langle\sigma^2_{\rm err}\rangle$ is the average photometric variance across the light curve. For the ZTF light curve, we find $\sigma_{\rm int}=0.039^{+0.003}_{-0.004}$ mag, and similarly for the CRTS light curve, we find $\sigma_{\rm int}=0.047^{+0.022}_{-0.047}$ mag, indicating weak variability over the combined ZTF and CRTS baseline. The errorbars on the intrinsic variability are simply the errorbars on the variance ${\rm var(y)}$, assuming that the ZTF or CRTS photometric errors $\sigma^2_{\rm err}$ are accurate. 

To perform an additional test for AGN-like variability, we use the Python package \texttt{eztao} which is built on top of the Gaussian process regression library \texttt{celerite} \citep{celerite} to fit the ZTF light curve of J1717+3807 to a Damped Random Walk (DRW; \citealt{macleod2010}), a well-known model for AGN variability. This model is characterized by an exponential covariance matrix $C_{\rm ij}=\sigma_{\rm DRW}^2 e^{-|t_i-t_j|/\tau_{\rm DRW}}$, where $\sigma_{\rm DRW}$ is the characteristic variability in units of magnitudes, $\tau_{\rm DRW}$ is a characteristic damping time scale, and $t_i, t_j$ are any two times on the light curve.  We find $\sigma_{\rm DRW}=0.043\; {\rm mag}, \tau_{\rm DRW}=0.04\; {\rm days}$ consistent with white noise. This is in contrast with typical quasars, which have longer damping timescales $\tau_{\rm DRW}$ of hundreds of days. To show this, we utilize the measured r-band DRW parameters from SDSS quasars \citep{macleod_2012}\footnote{\url{https://faculty.washington.edu/ivezic/macleod/qso_dr7/Southern.html}} and compare the DRW parameter distribution to DESI LRDs from this work in Fig.~\ref{fig:DRW_comparison} and Table~\ref{tab:ztf_variability}. To visualize the differences between AGN red-noise variability and (LRD) white-noise variability, for J1717 and BAQ1 we also calculate the power spectral density (PSD) of the light curve via a Lomb-Scargle (LS) periodogram \citep{Lomb1976,scargle1982}, fitting a line to the PSD $\propto f^{-\alpha}$ in log-log space to measure its slope. As can be seen in the bottom left panel of Fig.~\ref{fig:lc_analysis}, the best-fit slope $\alpha= 0.0036 \pm 0.008$ of this log-log line is consistent with 0, as expected for white noise.

Like J1717+3807, we find evidence for weak and uncorrelated white noise variability, based on the intrinsic scatter $\sigma_{\rm int}=0.0-0.1$ mag for most sources and small $\tau_{\rm DRW}$ for all. In particular for J1717+3807, J0716+5433, J0829+1312, this result holds for hundreds or thousands of observations. While J1502+0250 shows the most potential for significant variability, we are cautious to make such a conclusion given that most of its data points lie below the ZTF limiting magnitude. Moreover, the damping timescale $\tau_{\rm DRW}$ for J1502+0250 is much smaller than typical quasar variability (see also Fig.~\ref{fig:DRW_comparison}), indicative of white noise as also shown in the other DESI LRDs. ZTF r-band light curves of the DESI LRDs besides J1717+3807 can be seen in Figure~\ref{fig:lcs_new}. 

In contrast, we find that the light curve of BAQ1 (shown in top right panel of Fig.~\ref{fig:lc_analysis}) exhibits AGN-like variability. Measuring the DRW parameters in the same way using \texttt{eztao} gives $\sigma_{\rm DRW}=0.15\; {\rm mag}, \; \tau_{\rm DRW}=90.0\; {\rm days}$, which are typical of quasars \citep{macleod2010}. Identically calculating a Lomb-Scargle periodogram to estimate the PSD, we find a close match to the DRW PSD (pink-dashed curve in bottom right panel of Fig~\ref{fig:lc_analysis}), where \begin{equation}
    {\rm PSD}(f)=\frac{4\tau^2_{\rm DRW}\sigma^2_{\rm DRW}}{1+(2\pi f\tau_{\rm DRW})^2}. 
\end{equation} We also mark a characteristic "break" frequency $f_{\rm break}=1/(2\pi \tau_{\rm DRW})$, where the ${\rm PSD}\propto f^{-2}$ for red noise at $f>>f_{\rm break}$ but "flattens out" to white noise ${\rm PSD}\propto f^{0}$ at $f<<f_{\rm break}$.

\begin{table}
\caption{Variability statistics for the seven local LRD candidates and BAQ1 with available ZTF $r$-band light curves (J1137+5520 does not have ZTF data).}
\centering
\begin{tabular}{l l c c c c}
\toprule
 & Name & $N_{\rm obs, ZTF}$ & $\sigma_{\rm int}$ & $\sigma_{\rm DRW}$ & $\tau_{\rm DRW}$ \\
\midrule
 \vspace{0.05cm}
 & J1717+3807 & 2061 & $0.04^{+0.003}_{-0.004}$ & 0.04  & $8\times 10^{-5}$  \\
 \vspace{0.05cm}
 & J0716+5433 & 346 & $0.00^{+0.04}_{-0.00}$ & 0.14  & 3.0  \\
  \vspace{0.05cm}
 & J0829+1312 & 252 & $0.11^{+0.03}_{-0.04}$ & 0.20 & 2.7  \\
  \vspace{0.05cm}
 & J1502+0250 & 97 & $0.23^{+0.05}_{-0.07}$ & 0.32 & 5.4 \\
  \vspace{0.05cm}
 & J1343+3934 & 81  & $0.11^{+0.06}_{-0.11}$ & 0.19 & 0.03 \\
  \vspace{0.05cm}
 & J0129+0628 & 14  & $0.05^{+0.08}_{-0.05}$ & 0.09 & $7.2\times10^{-5}$ \\
  \vspace{0.05cm}
 & J1909+5831 & 3  & $0.54^{+0.10}_{-0.54}$ & N/A & N/A\\
  \vspace{0.05cm}
 & J1137+5520 & 0  & N/A & N/A & N/A\\ \hline
 \vspace{0.05cm}
 & BAQ1 & 917 & $0.14^{0.003}_{-0.003}$ & 0.15 & 90.0\\
\bottomrule
\end{tabular}
\tablefoot{We display in order of the number of data points of the ZTF $r$ band light curve. $\sigma_{\rm int}$ is the intrinsic variability in units of magnitude, $\sigma_{\rm DRW}, \tau_{\rm DRW}$ are the AGN Damped Random Walk parameters in units of magnitude and days, respectively. We do not attempt to calculate DRW parameters for J1909+5831, which has a small number of data points.}
\label{tab:ztf_variability}
\end{table}

The light curves of the J1717+3801 and BAQ1 are shown as an illustrative example that indicates that the variability is correlated with the overall spectral shape of the sources. If intrinsic variability in the engines of LRDs were to be regulated by dense gas envelopes that is seen in the Balmer absorption lines \citep[e.g.,][]{kido2025}, such mechanism is clearly not present in BAQ1, again suggesting that the covering fraction of the absorber is much smaller. It could therefore be promising to use the lack of (strong) variability as a selection criterion to identify local LRDs. We expect that more rigorous statistical analyses of variability will be enabled with the discovery of more local LRDs and the capability of LSST to monitor faint sources down to $24-26$ mag as well as further extending the existing baselines to decades.

\section{Discussion }
\label{sec:discussion}

\subsection{Number Density of low-redshift LRDs} \label{sec:numberdens}

\begin{figure}
    \centering
    \includegraphics[width=\linewidth]{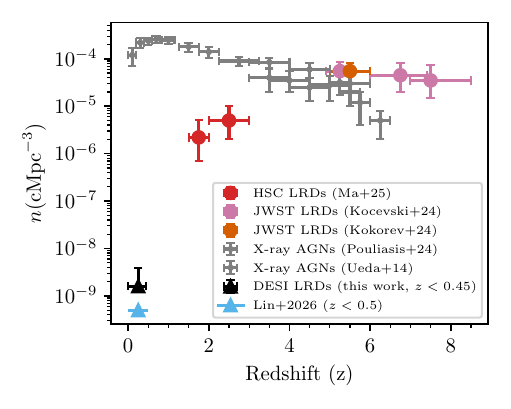}
    \caption{Number density estimate of DESI LRDs compared to the estimate of \cite{lin2026}, high-redshift estimates \citep{kocevski2024, kokorev2024a, ma2025_counting}, and the numbers of X-ray AGNs \citep{ueda_x_ray_agn, Pouliasis_agn}. The upper errorbar on the DESI estimate assumes that all ambiguous sources in our search are LRDs, although this is highly unlikely. }
    \label{fig:number_densities}
\end{figure}
The evolution of the number density of LRDs is one of the key observational tests for models of the nature of LRDs. Given that we have identified eight LRD candidates, it is possible to make a crude estimate of the number density of LRDs at low redshift. The comoving volume within $z<0.45$ is $V_{z<0.45}\sim21.4\; {\rm Gpc^3}$ and the DESI survey footprint for DR1 is $A_{\rm DESI}\sim 9528\; {\rm deg^2}$ \citep{DESI_DR1}. If we assume $f_{\rm complete}$ as a fudge factor representing the fraction of galaxies that DESI has observed in DR1 in its footprint out of the true number of galaxies, and we assume these 8 sources are genuine local LRDs, we can estimate the number density of LRDs as 
$n_{\rm LRD}(z<0.45)\times V_{z<0.45}\frac{A_{\rm DESI}}{A_{\rm sky}}f_{\rm complete}\sim8$. Assuming $A_{\rm sky}=41253\; {\rm deg^2}$ gives \begin{equation}
    n_{\rm LRD}(z<0.45)= f_{\rm complete}^{-1}\times1.6\times 10^{-9}\; {\rm Mpc^{-3}}.
\end{equation} We note that a similar estimate made for the LRD number density based on SDSS data \citep{lin2026} yielded a comparable estimate $n_{\rm LRD}(z<0.5)\gtrsim 5\times 10^{-10} \; {\rm Mpc^{-3}}$. The number densities would be higher (see upper errorbar in Fig.~\ref{fig:number_densities}) if all our {\it ambiguous} sources were to be confirmed as LRDs, although this is unlikely as discussed in Section $\ref{sec:Methods}$. The spectroscopic completeness is the key uncertainty. It is very challenging to reliably estimate $f_{\rm complete}$ because the identified LRDs span a range of selections and extend to the faintest magnitudes in DESI (see \citealt{Myers23} for a detailed overview of the complex aspects of the DESI target selection). Most LRDs were part of the quasi-stellar object target class due to their compactness. J1717+3807 and J0829+1312 were also selected in the Luminous Red Galaxy target class. J1909+5831, however, was only observed as a filler target. The least that we can say is that our results yield a firm lower limit to the number density by assuming $f_{\rm complete}=1$.

\subsection{Comparison to \cite{Ding2026}}

\cite{Ding2026} recently also performed a search for (analogues of) LRDs in DESI DR1. Unlike our spectroscopic approach, their target selection mimicks the photometric approaches used in JWST data \citep[e.g.,][]{kokorev2024a} that use a compactness criterion as well as a V-shaped color based on UV ($\lambda<3645 \AA$) and optical ($\lambda>3645 \AA$) SED slopes. There is no overlap in our samples. This is because we remove sources with strong [\ion{N}{ii}] and [\ion{S}{ii}] lines as those are not seen among high-redshift LRDs, whereas they are common among Seyferts and normal quasars. The lack of strong [\ion{N}{ii}] and [\ion{S}{ii}] in LRDs is attributed to their high covering fractions of gas with densities well above the critical densities of these lines, as well as their low gas-phase metallicity. As photoionization models suggest that this dense gas is the cause of the V-shaped spectrum of LRDs \citep[e.g.,][]{inayoshi2025,ji2025,naidu2025}, the detection of strong [\ion{N}{ii}] and [\ion{S}{ii}] in the sources studied by \cite{Ding2026} suggests that their V-shaped SEDs are likely due to other physical mechanisms.

\subsection{A Dearth of Red and Luminous LRDs at $z\sim 0$?} \label{sec:dearth}
\begin{figure*}
    \centering\includegraphics[width=0.49\linewidth]{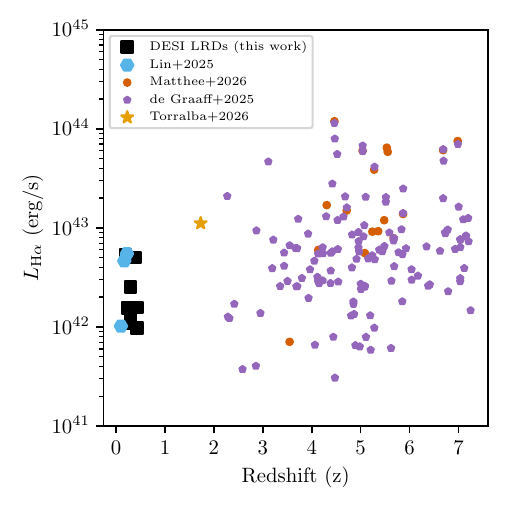}
    \includegraphics[width=0.49\linewidth]{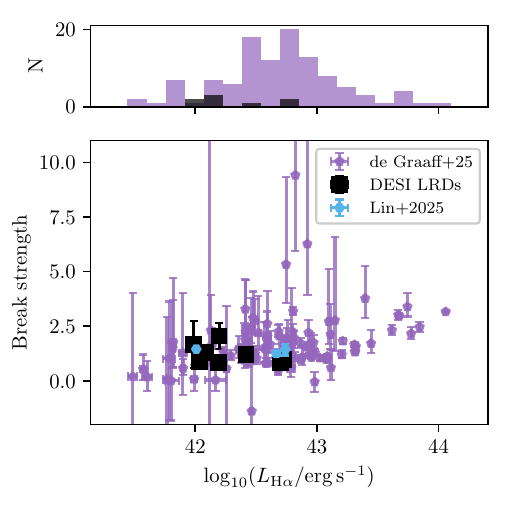}
    \caption{Left panel: Distribution of H$\alpha$ luminosity as a function of redshift for DESI local LRDs (this work), and other known LRDs \cite{lin2026, Matthee2026, degraaff2025, torralba2026_bh_star}. Right panel: Balmer break strength vs H$\alpha$ luminosity for the \cite{degraaff2025} sample compared to DESI local LRDs.}
    \label{fig:L_ha_vs_z}
\end{figure*}

As we showed in our stacked spectrum (Section $\ref{sec:stacks}$), the DESI LRDs most resemble the high-redshift LRDs with relatively weak Balmer breaks. Fig.~\ref{fig:L_ha_vs_z} further elaborates these differences by showing the distribution of H$\alpha$ luminosities and Balmer break strengths of the low-redshift LRDs as well as a large sample at high-redshift.
We have computed the Balmer break strength following \citet{degraaff2025}, as the median ratio $f_{\nu, 4000-4100}/f_{\nu, 3620-3720}$. We note that some of our candidates have a Balmer break strenght $<1$. For instance, the LRD J1717+3807 highlighted in Fig.~\ref{fig:J1717_spectrum} presents a clear Balmer jump, which could indicate a significant contribution from a young star-forming galaxy to the SED of this object \citep[see][for a discussion over a highly similar LRD spectrum with a Balmer jump]{Killi24}. Alternatively, Balmer jumps could also be produced in dense gaseous envelopes of LRDs under specific conditions, as predicted by models \citep[e.g.,][]{sneppen26, Liu2026_atmosphere}, possibly due to a higher covering fraction of lower column density channels \citep[e.g.][]{Tang26}.
The H$\alpha$ luminosities in our sample are in the range of $(1-5)\times 10^{42}{\rm erg/s}$, similar to the sample from \cite{lin2025}. Thus, compared to high-redshift LRDs \citep{Matthee2026, degraaff2025}, the low-redshift LRDs lack the most luminous LRDs with $L_{\rm H\alpha}>10^{43}$~\unit{erg.s^{-1}}. 

The lack of red and luminous LRDs at $z\sim0$ could either be due to selection effects, or physical evolution. As shown in the right panel of Fig.~\ref{fig:seds8}, it appears that more luminous LRDs are redder, especially those above L$_{\rm H\alpha}>10^{43}$ erg s$^{-1}$. Given that most LRDs among our sample were DESI-selected as QSO targets and some as possible LRGs, it is hard to imagine how the most luminous sources at low-redshift would be missed. Nevertheless, blind wide-area spectroscopic surveys, for example with Euclid or Roman would be very valuable in addressing such possible biases introduced by photometric pre-selection. 

If physical effects would cause the lack of luminous LRDs at low-redshift, we speculate that this could be due to the larger metallacities in low-redshift environments leading to efficient cooling and to lower gas supplies available for rapid collapse of gas fueling the most luminous LRDs at lower redshifts. Given that the redness and high H$\alpha$ EWs in LRDs have been associated with high gas accretion rates \citep[e.g.,][]{madau_maiolino_2026}, the lack of H$\alpha$-luminous and red LRDs at low-redshifts, and their lower H$\alpha$ EWs compared to the high-redshift sources, aligns with such a picture of decreased gas accretion rates.

\section{Summary}\label{sec:summary}

Samples of Little Red Dots at low redshift offer valuable insights into their nature as they are suitable for sensitive follow-up across the electromagnetic spectrum. Moreover, their lower redshift facilitates variability studies and the redshift evolution of their number density and luminosity function provides important insights into the physical conditions that give rise to the phenomenon. We performed a spectroscopic search for low-redshift Little Red Dots by comparing the rest-frame optical spectra of broad-emission line sources in the DESI DR1 catalog with templates based on high-redshift observations. Additionally, we remove objects with H$\alpha$ lines with relatively low equivalent width and line-width, as well as quasars with high-excitation [NeV] line emission (see Table. $\ref{tab:selection_criteria}$). Here, we summarize our results from this search:

\begin{itemize}
    \item We discover eight low-redshift LRDs at redshifts $z=0.20-0.44$  which show properties of high-redshift LRDs: blue-shifted, narrow absorption in the H$\alpha$ line in 5/8 cases, broad exponential wings, strong He I line-emission that anti-correlates with the (steep) Balmer decrements, weak IR emission and possible inflection points in the SED near the Balmer break wavelength. While it was not required as a search criterion, the 8 LRDs show compact morphologies that are unresolved in ground-based imaging with $\sim1\arcsec$ resolution. [Table $\ref{tab:basic_properties}$, Figs. $\ref{fig:J1717_spectrum}$, $\ref{fig:ha_all}$, $\ref{fig:HaHbratio_and_HeI}$ and $\ref{fig:seds8}$]

    \item Compared to high-redshift LRDs, our sources show similar strengths in the narrow (forbidden) lines as [\ion{Fe}{ii}], \ion{He}{i}, [\ion{Ar}{iv}], and [\ion{S}{iii}], but the Balmer line equivalent widths are a factor $\sim1.5$ lower. Nebular HeII emission is detected, but its line-ratio with respect to H$\beta$ is much lower than typical Type I AGNs, similar to high-redshift LRDs and suggesting a cut-off in the spectrum of the ionising source around 50 eV. [Section $\ref{sec:stacks}$, Fig. $\ref{fig:stacks_full}$]

    \item Photometric time-domain data is available with a relatively high cadence from the ZTF and CRTS surveys. The LRDs show weak $0.0-0.1$ mag variability across multiple-year-long baselines in their rest-frame, and are consistent with white noise. Quantifying the variability with a damped random walk model, we show that the variability of DESI LRDs is unusual compared to AGN red-noise variability with damping timescales of hundreds of days.  [Section $\ref{sec:variability}$, Figs. $\ref{fig:DRW_comparison}$ and $\ref{fig:lc_analysis}$]

    \item During our search, we found a quasar which we nickname "Balmer Absorption Quasar 1" (BAQ1), which shows very similar Balmer line-profiles to LRDs, including narrow blue-shifted absorption, yet it shows typical quasar features as high-excitation lines and broad permitted FeII. BAQ1 shows clear variability with the typical AGN red-noise power spectral density. These results highlight the potential of using the (lack of) variability as a highly specific way of selecting LRDs at low redshifts.  [Section $\ref{sec:balmer_absorption_qsos}$, Figs. $\ref{fig:baq1_balmer}$ and $\ref{fig:DRW_comparison}$] 

    \item Our search implies a number density of $f^{-1}_{\rm complete} 1.6\times 10^{-9} \; {\rm Mpc^{-3}}$, where $f_{\rm complete}$ is the spectroscopic completeness of the DESI DR1 survey. Unless $f_{\rm complete}$ is very small, our results imply orders of magnitude lower number densities of LRDs in the low-redshift Universe, compared to $z>2$, similar to earlier estimates based on the SDSS. Combined with other samples, we report a lack of red LRDs with luminous H$\alpha$ lines at $z\sim0$. While a selection effect cannot be ruled out and such luminous sources should easily be detectable in blind wide-area spectroscopic surveys, we speculate that this could be related to differences in the gas-phase metallicity that controls the gas mass available for rapid collapse to fuel LRD activity. [Sections $\ref{sec:numberdens}$ and $\ref{sec:dearth}$,  Figs. $\ref{fig:number_densities}$ and $\ref{fig:L_ha_vs_z}$]

\end{itemize}

\section*{Acknowledgements}

We thank Aayush Desai for useful conversations on how to perform this search efficiently. 

Funded by the European Union (ERC, AGENTS,  101076224). Views and opinions expressed are however those of the author(s) only and do not necessarily reflect those of the European Union or the European Research Council. Neither the European Union nor the granting authority can be held responsible for them. 

We acknowledge support by NASA grants 80NSSC22K0822 and 80NSSC24K0440 (ZH). 

This research uses services or data provided by the Astro Data Lab, which is part of the Community Science and Data Center (CSDC) Program of NSF NOIRLab. NOIRLab is operated by the Association of Universities for Research in Astronomy (AURA), Inc. under a cooperative agreement with the U.S. National Science Foundation. In addition, this research uses services or data provided by the SPectra Analysis and Retrievable Catalog Lab (SPARCL) and the Astro Data Lab, which are both part of the Community Science and Data Center (CSDC) Program of NSF NOIRLab. NOIRLab is operated by the Association of Universities for Research in Astronomy (AURA), Inc. under a cooperative agreement with the U.S. National Science Foundation. 

Based on observations obtained with the Samuel Oschin Telescope 48-inch and the 60-inch Telescope at the Palomar Observatory as part of the Zwicky Transient Facility project. ZTF is supported by the National Science Foundation under Grant No. AST-2034437 and a collaboration including Caltech, IPAC, the Weizmann Institute for Science, the Oskar Klein Center at Stockholm University, the University of Maryland, Deutsches Elektronen-Synchrotron and Humboldt University, the TANGO Consortium of Taiwan, the University of Wisconsin at Milwaukee, Trinity College Dublin, Lawrence Livermore National Laboratories, and IN2P3, France. Operations are conducted by COO, IPAC, and UW.

This research used data obtained with the Dark Energy Spectroscopic Instrument (DESI). DESI construction and operations is managed by the Lawrence Berkeley National Laboratory. This material is based upon work supported by the U.S. Department of Energy, Office of Science, Office of High-Energy Physics, under Contract No. DE–AC02–05CH11231, and by the National Energy Research Scientific Computing Center, a DOE Office of Science User Facility under the same contract. Additional support for DESI was provided by the U.S. National Science Foundation (NSF), Division of Astronomical Sciences under Contract No. AST-0950945 to the NSF’s National Optical-Infrared Astronomy Research Laboratory; the Science and Technology Facilities Council of the United Kingdom; the Gordon and Betty Moore Foundation; the Heising-Simons Foundation; the French Alternative Energies and Atomic Energy Commission (CEA); the National Council of Humanities, Science and Technology of Mexico (CONAHCYT); the Ministry of Science and Innovation of Spain (MICINN), and by the DESI Member Institutions: www.desi.lbl.gov/collaborating-institutions. The DESI collaboration is honored to be permitted to conduct scientific research on I’oligam Du’ag (Kitt Peak), a mountain with particular significance to the Tohono O’odham Nation. Any opinions, findings, and conclusions or recommendations expressed in this material are those of the author(s) and do not necessarily reflect the views of the U.S. National Science Foundation, the U.S. Department of Energy, or any of the listed funding agencies.

{\it Software:} {\tt python} \citep{travis2007,jarrod2011}, {\tt numpy} \citep{walt2011}, {\tt matplotlib} \citep{hunter2007}, {\tt scipy} \citep{2020SciPy}, {\tt astropy} \citep{astropy:2013, astropy:2018, astropy:2022}, {\tt astroML} \citep{astroMLText}, {\tt datalab}, {\tt sparcl} \citep{datalab1, datalab2, datalab3}


\appendix
\section{LRD Spectra and Ambiguous Sources}

\begin{figure*}[h]
    \centering
    \includegraphics[width=\linewidth]{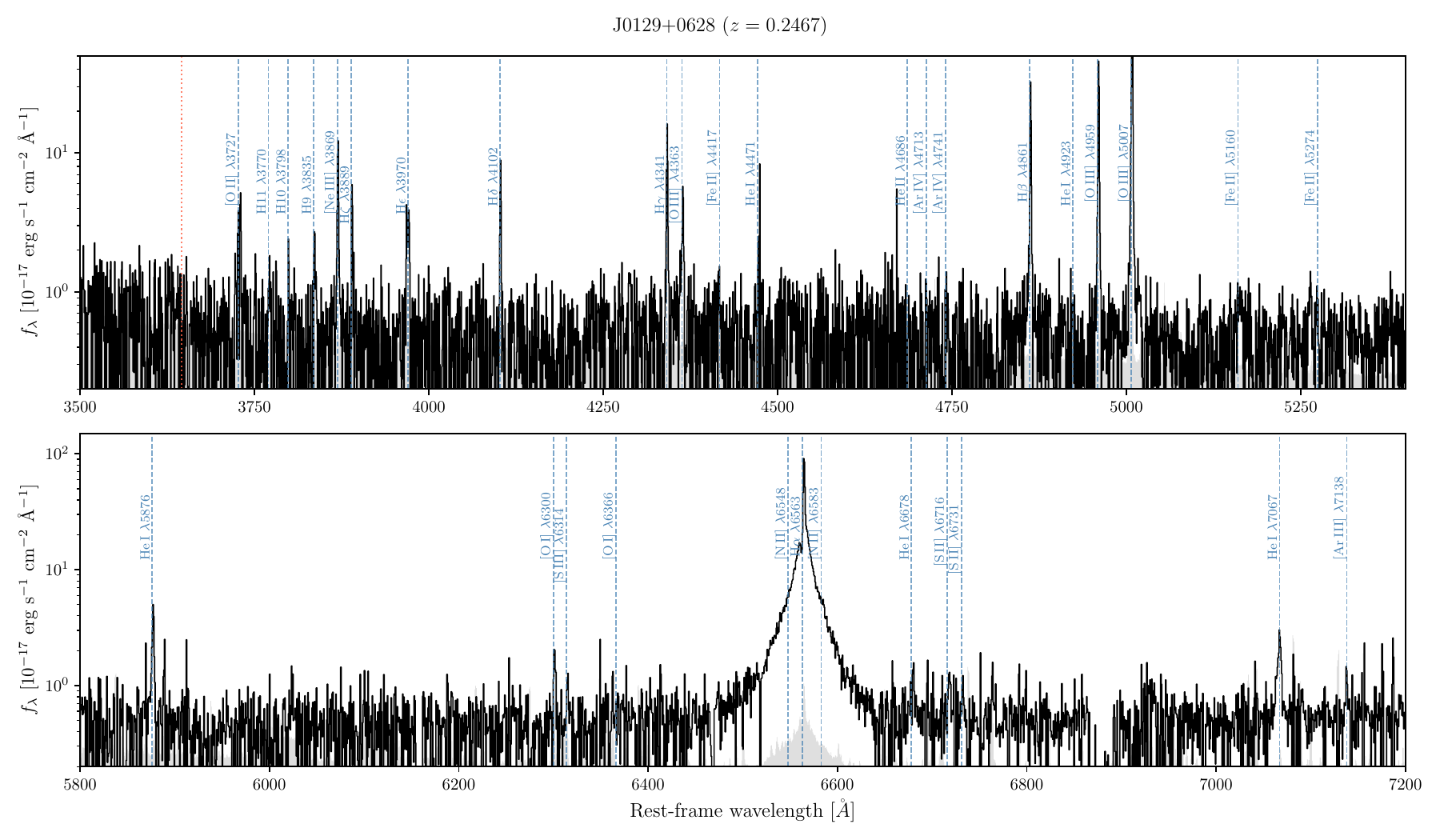}
    \includegraphics[width=\linewidth]{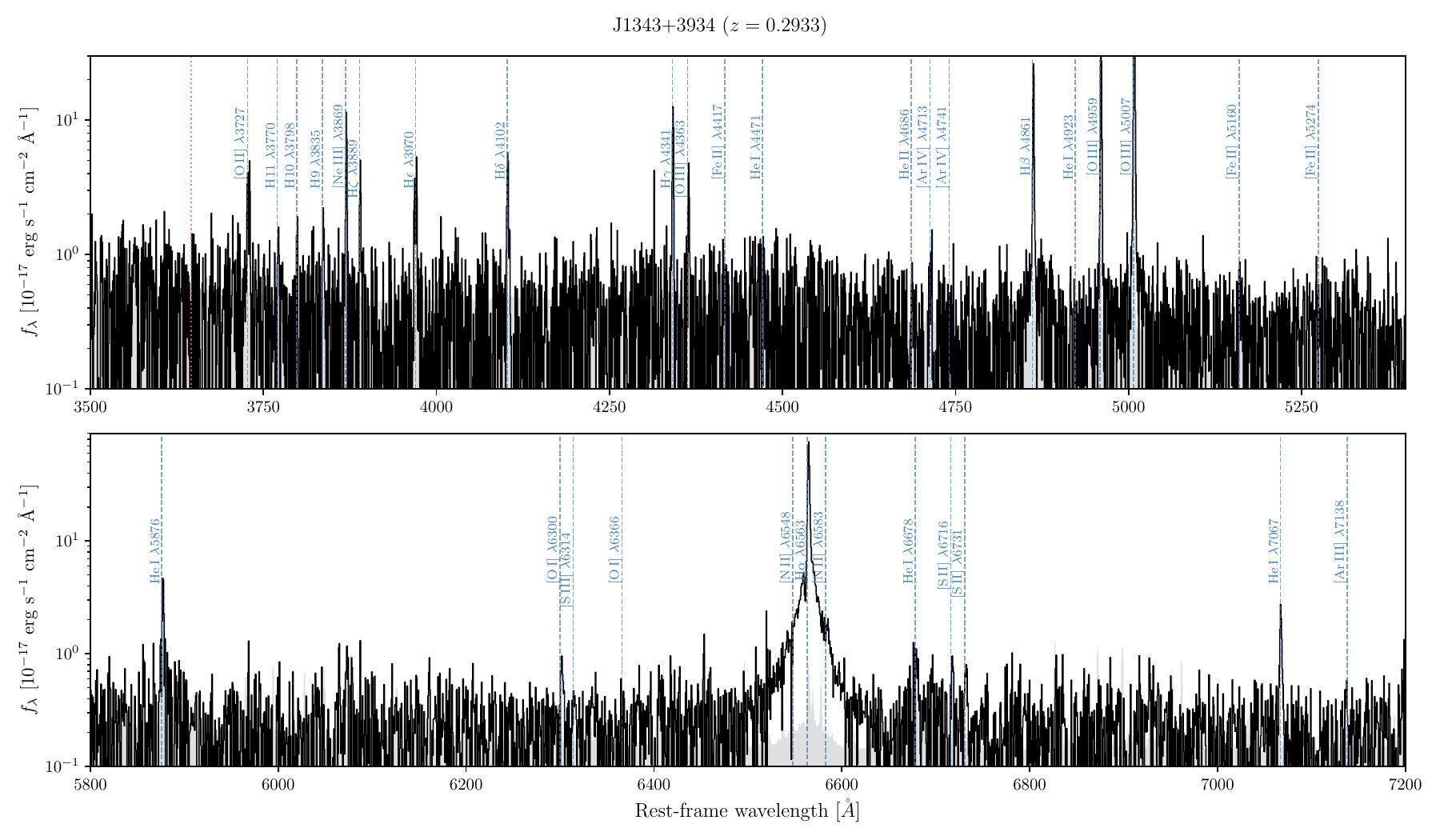}
    \caption{DESI spectra of LRDs J0129+0628 and J1343+3934 identified in this work (as in Fig.~\ref{fig:J1717_spectrum}).}
    \label{fig:J0129_J1343_full}
\end{figure*}

\begin{figure*}[h]
    \centering
    \includegraphics[width=\linewidth]{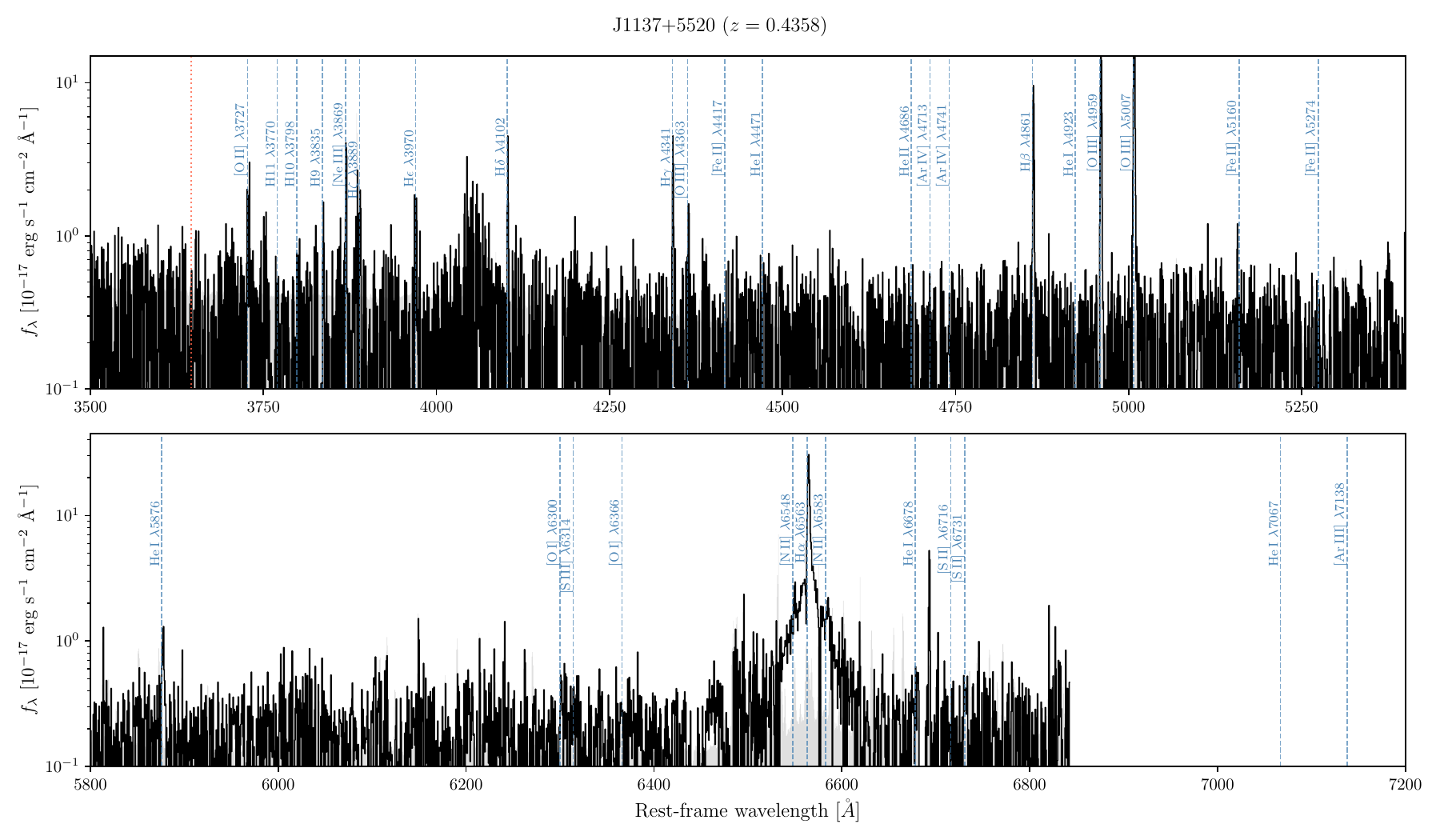}
    \includegraphics[width=\linewidth]{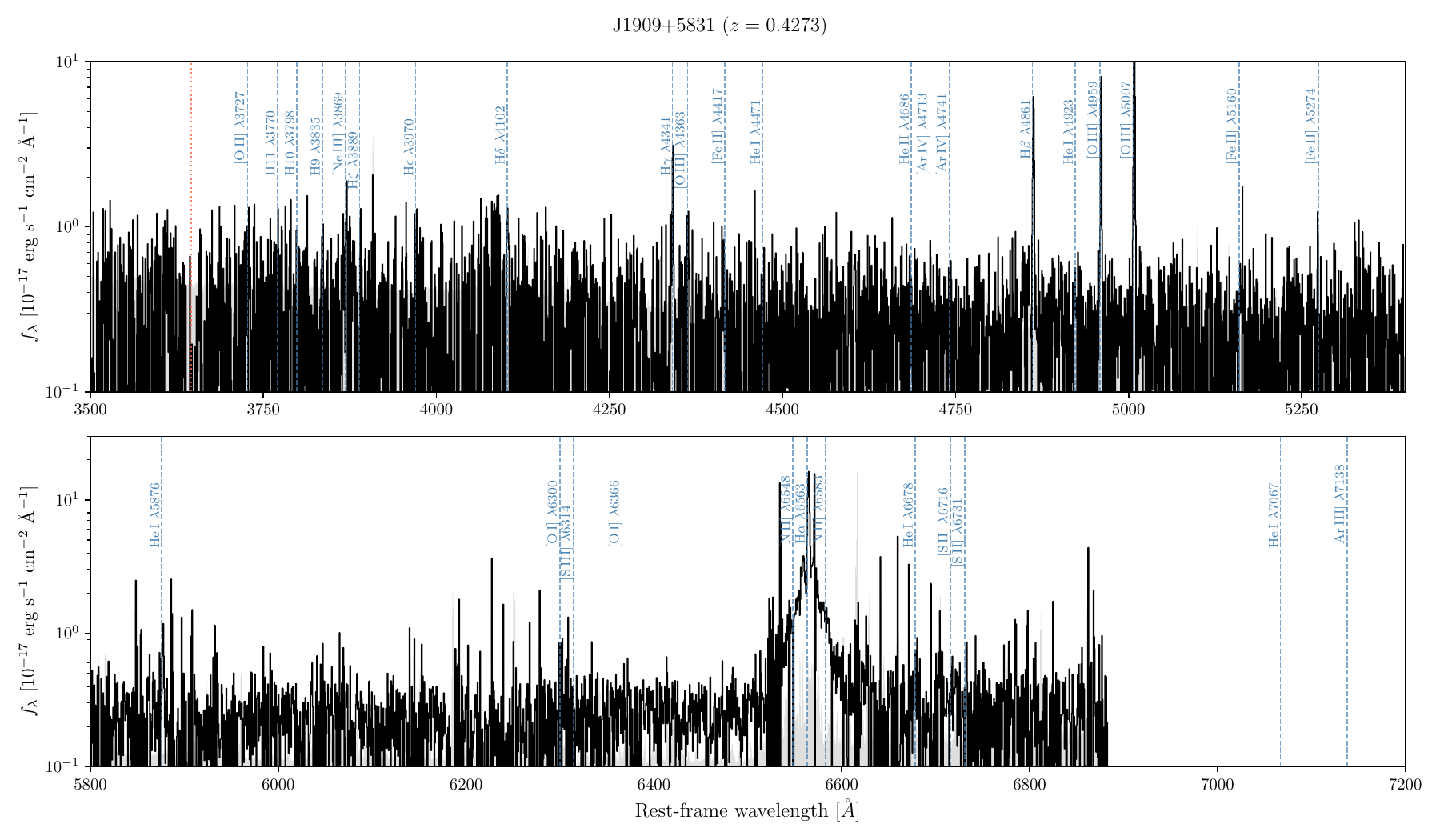}
    \caption{DESI spectra of LRDs J1137+5520 and J1909+5831 identified in this work (as in Fig.~\ref{fig:J1717_spectrum}).}
    \label{fig:J1137_J1909_full}
\end{figure*}

\begin{figure*}[h]
    \centering
    \includegraphics[width=\linewidth]{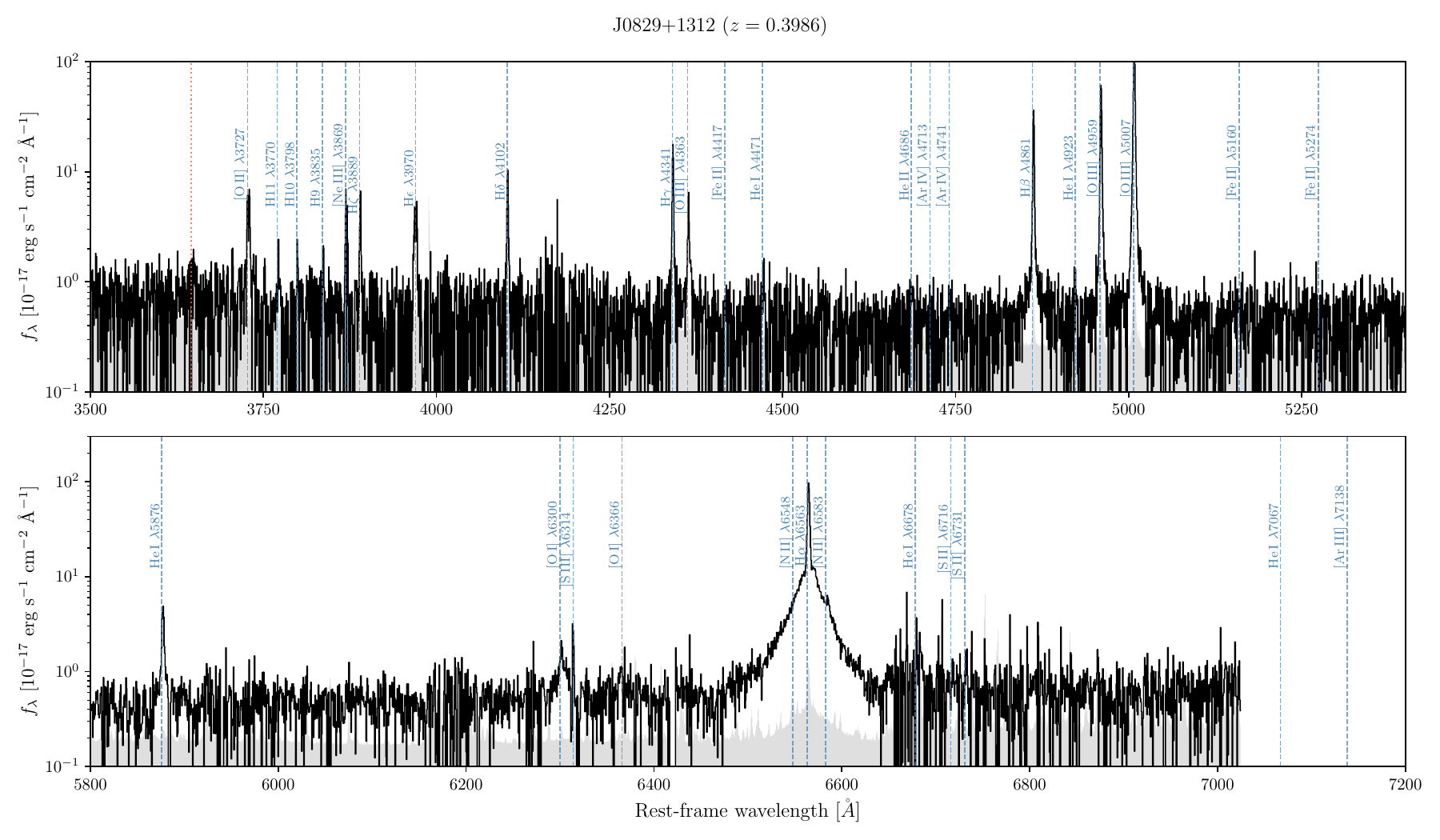}
    \includegraphics[width=\linewidth]{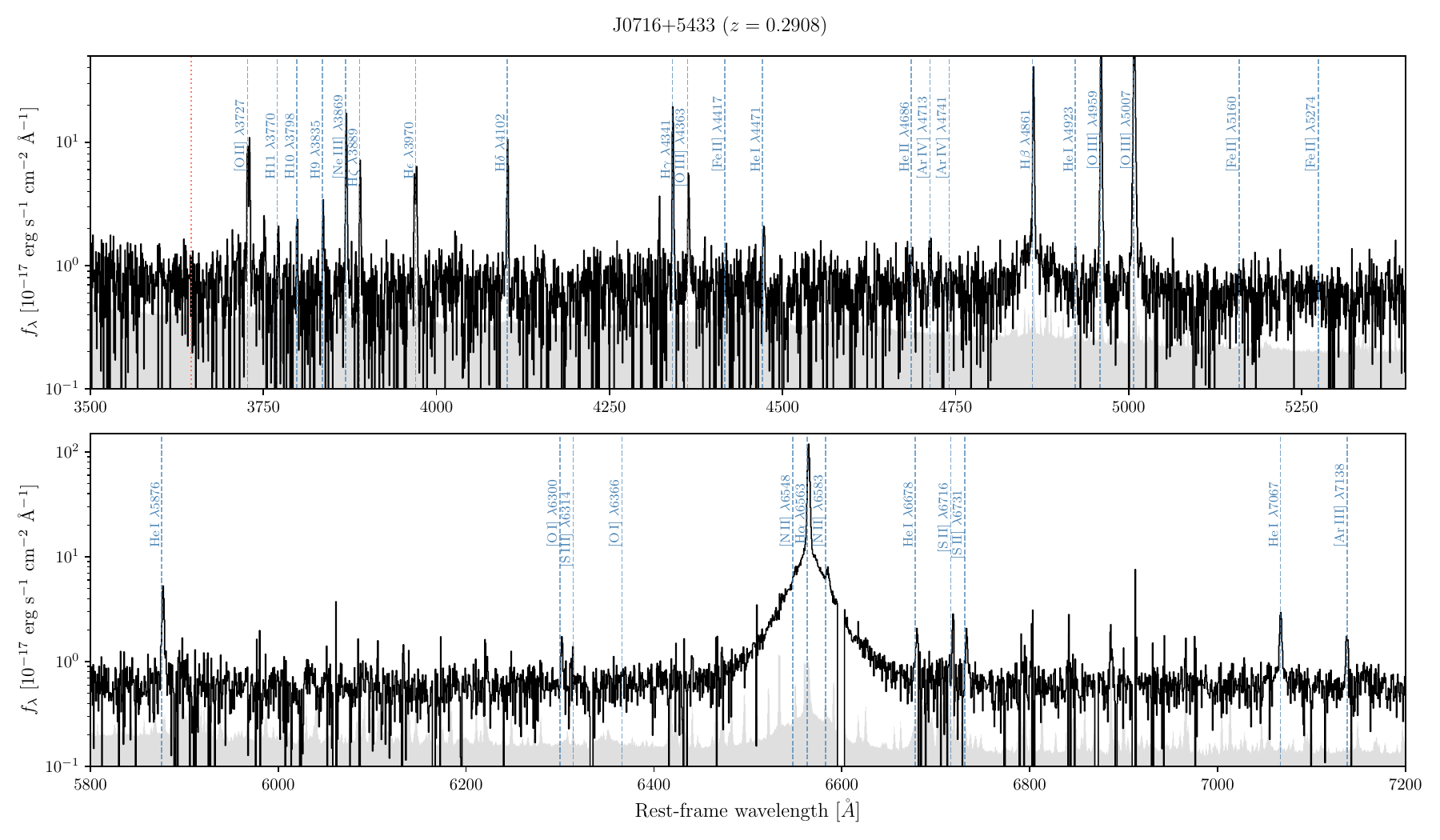}
    \caption{DESI spectra of LRDs J0829+1312 and J0716+5433 identified in this work (as in Fig.~\ref{fig:J1717_spectrum}).}
    \label{fig:J0829_J0716_full}
\end{figure*}

\begin{figure*}[h]
    \centering
    \includegraphics[width=\linewidth]{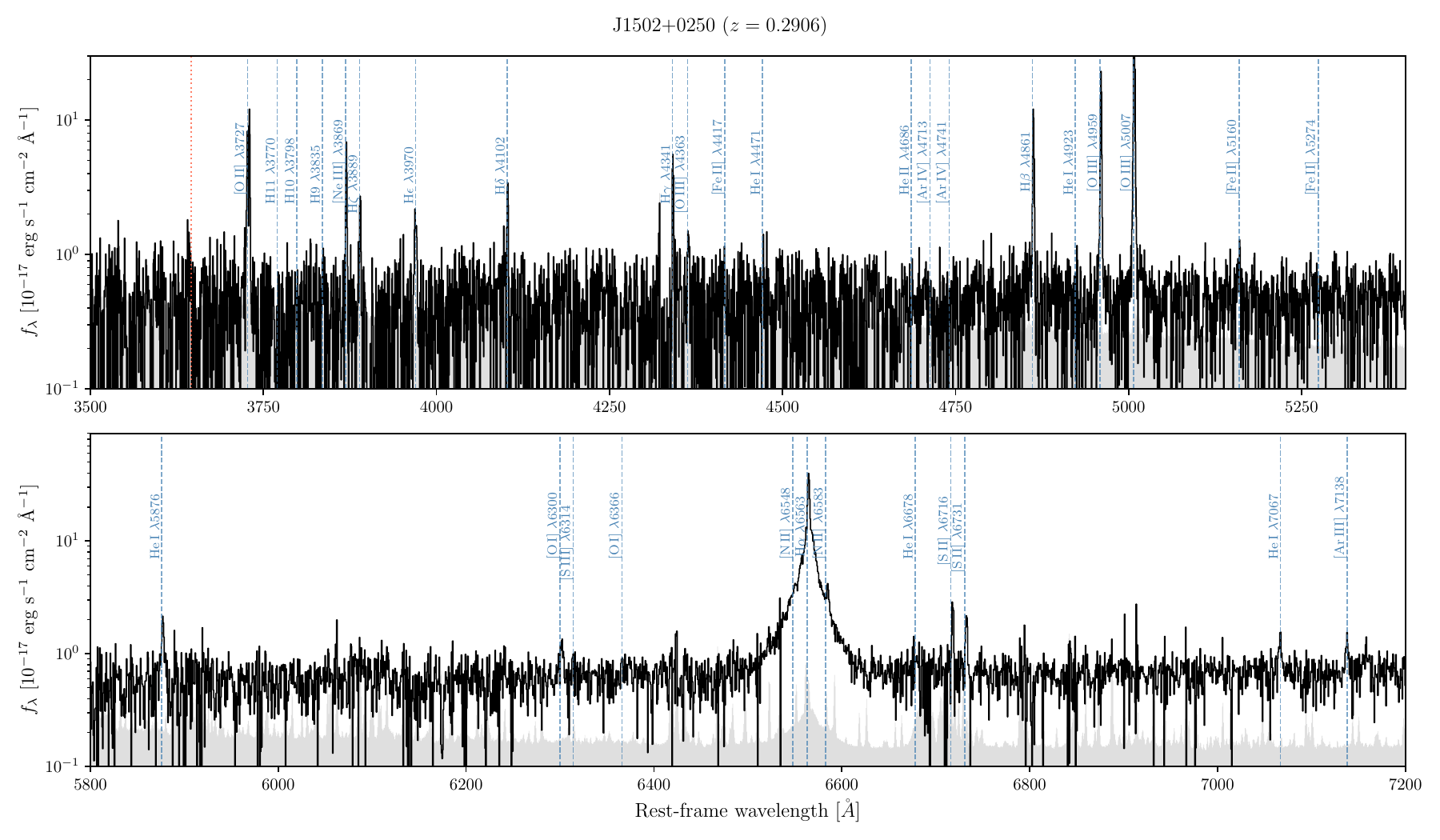}
    \caption{DESI spectra of LRD J1502+0250 identified in this work (as in Fig.~\ref{fig:J1717_spectrum}).}
    \label{fig:J1502_full}
\end{figure*}

In Fig.~\ref{fig:J0129_J1343_full}, Fig.~\ref{fig:J1137_J1909_full}, Fig.~\ref{fig:J0829_J0716_full}, and Fig.~\ref{fig:J1502_full}, we show the DESI spectra of the LRDs identified in this work excluding J1717+3807 (shown in Fig.~\ref{fig:J1717_spectrum}). As in Fig.~\ref{fig:J1717_spectrum}, we plot the wavelength ranges of $3500-5400 \AA$ and $5800-7200 \AA$, which include relevant spectral lines in our analysis.

In Table~\ref{tab:ambiguous1}, we show coordinates and DESI IDs for the final ambiguous sources which we could not rule out from our search criteria.

\begin{table}[h]
\caption{DESI spectra viewer, DESI target id, coordinates (in J2000) and redshifts of the final 11 ambiguous sources from our systematic search (see Table~\ref{tab:selection_criteria} for details). }
\centering
\begin{tabular}{rrrc}
\toprule
DESI targetid & R.A. & Dec. & $z$ \\
\midrule
\href{https://www.legacysurvey.org/viewer/desi-spectrum/dr1/targetid39627515605554125}{39627515605554125} & 66.3673 & -11.1843 & 0.1577 \\
\href{https://www.legacysurvey.org/viewer/desi-spectrum/dr1/targetid39627731482186838}{39627731482186838} & 66.4798 & -2.2700 & 0.1852 \\
\href{https://www.legacysurvey.org/viewer/desi-spectrum/dr1/targetid39627782354897895}{39627782354897895} & 218.6236 & -0.3037 & 0.4483 \\
\href{https://www.legacysurvey.org/viewer/desi-spectrum/dr1/targetid39627889422894267}{39627889422894267} & 122.4551 & 4.1357 & 0.4350 \\
\href{https://www.legacysurvey.org/viewer/desi-spectrum/dr1/targetid39627908985128696}{39627908985128696} & 211.4203 & 4.9916 & 0.4317 \\
\href{https://www.legacysurvey.org/viewer/desi-spectrum/dr1/targetid39627911430409814}{39627911430409814} & 357.6689 & 5.1077 & 0.4334 \\
\href{https://www.legacysurvey.org/viewer/desi-spectrum/dr1/targetid39628023418329441}{39628023418329441} & 245.1517 & 9.6297 & 0.4086 \\
\href{https://www.legacysurvey.org/viewer/desi-spectrum/dr1/targetid39628225134985394}{39628225134985394} & 37.6389 & 18.4696 & 0.3287 \\
\href{https://www.legacysurvey.org/viewer/desi-spectrum/dr1/targetid39628461970556460}{39628461970556460} & 336.9661 & 29.0997 & 0.0035 \\
\href{https://www.legacysurvey.org/viewer/desi-spectrum/dr1/targetid39628518019040271}{39628518019040271} & 245.1347 & 31.8047 & 0.4099 \\
\href{https://www.legacysurvey.org/viewer/desi-spectrum/dr1/targetid39633282664497740}{39633282664497740} & 214.5969 & 52.1818 & 0.2071 \\
\bottomrule
\end{tabular}
\label{tab:ambiguous1}
\end{table}

\section{Light Curves} 

\begin{figure*}
    \caption{ZTF r-band light curves of the local LRDs. J1717+3801 is omitted because it is shown in Fig.~\ref{fig:lc_analysis} and J1137+5520 does not have any observations to our knowledge.}
    \centering
    \includegraphics[width=\linewidth]{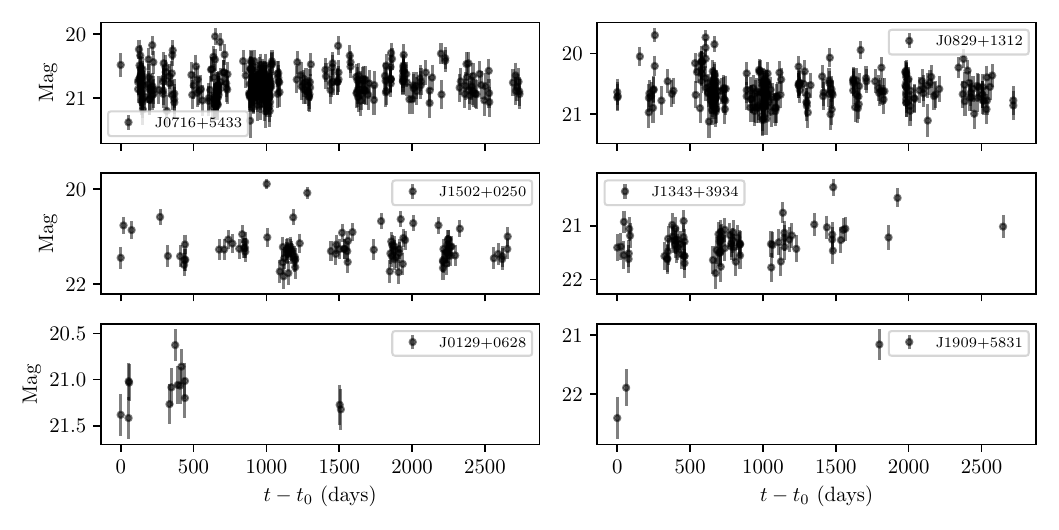}
    \label{fig:lcs_new}
\end{figure*}

We also show the ZTF r-band light curves of the DESI local LRDs in Fig.~\ref{fig:lcs_new}. The analysis of these light curves can be seen in Section~\ref{sec:variability}.

\section{Balmer line properties of BAQs}

We present absorption properties in the Balmer lines (H$\alpha$-H$\epsilon$) of Balmer Absorption Quasar (BAQ) 1 in Table.~\ref{tab:BAQ1_absorber}. As mentioned in Section~\ref{sec:balmer_absorption_qsos}, the absorption features are similar to high-z LRDs, but BAQ1 shows prominent [NeV] and HeII along with variability in its optical light curve, which are different compared to LRDs.

\begin{figure*}
    \centering
    \includegraphics[width=\linewidth]{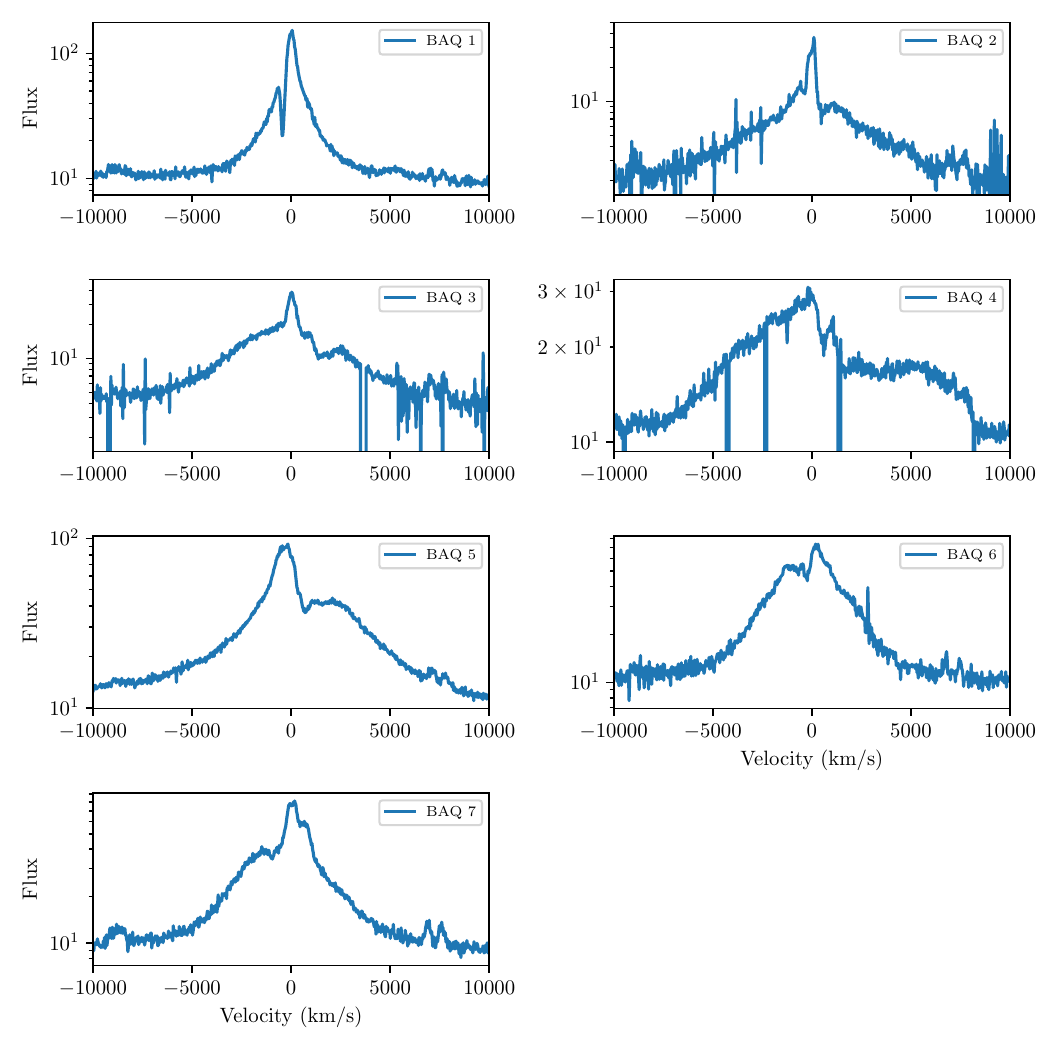}
    \caption{H$\alpha$ spectra of 7 quasars with apparent Balmer absorption features in their spectra. The profiles show a wide variety, in some case indicative of broad Keplerian disks combined with lines with intermediate widths. BAQ 1 is the only Balmer absorption-line quasar that we identified whose profile resembles the LRD profiles.  }
    \label{fig:baq_has}
\end{figure*}

We also show the H$\alpha$ profiles of other quasars with absorption features in Fig.~\ref{fig:baq_has}. The absorption features in these sources are both redshifted and blueshifted but only BAQ 1 shows sharp and narrow absorption features like LRDs.

\begin{table}[h]
\centering
\begin{tabular}{ccc}
\hline
Line & Vel. Offset [km\,s$^{-1}$] & Abs. Depth $\tau_0$ \\
\hline
 H$\alpha$ & $-369.9 \pm 4.6$  & $1.8 \pm 0.1$ \\
H$\beta$ & $-352.8 \pm 17.9$ & $4.9 \pm 2.7$ \\
H$\gamma$ & $-500.2 \pm 37.7 $& $124.8 \pm 3273.4$* \\
H$\delta$ & $-369.2 \pm 69.7$ &  $8.2 \pm 11.0$\\
H$\epsilon$ & $-309.7 \pm 2166.7$ & $11.8 \pm 535.7$ \\
\hline
\end{tabular}
\caption{Properties of the absorber of BAQ1 present in its Balmer lines. * We note that the large errors and extremely large absorption depths for H$\gamma$, H$\delta$, H$\epsilon$ reflect the fact that the absorption in these lines are close to or even slightly below the continuum, which our model (see Eq.~\ref{eq:Ha_model}) cannot produce and which suggests that (unlike in LRDs) the absorbing gas is located outside of the region that produces the continuum emission. }
\label{tab:BAQ1_absorber}
\end{table}

\bibliographystyle{aa}
\bibliography{my_bibliography}

\end{document}